\documentclass[twocolumn,twocolappendix,trackchanges]{aastex63}
\usepackage{amsmath}
\usepackage{showyourwork}
\newcommand{\code}[1]{\texttt{#1}}
\newcommand{\mesa}{\code{MESA}}

\usepackage{CJK}
\DeclareRobustCommand{\Eqref}[1]{Eq.~\ref{#1}}
\DeclareRobustCommand{\Figref}[1]{Fig.~\ref{#1}}

\DeclareRobustCommand{\Secref}[1]{Sec.~\ref{#1}}

\begin{document}

\graphicspath{{./figures/}}

\title{Rejuvenated accretors have less bound envelopes:\\ Impact of
  Roche lobe overflow on subsequent common envelope events}

\author[0000-0002-6718-9472]{M.~Renzo}
\affiliation{Center for Computational Astrophysics, Flatiron Institute, New York, NY 10010, USA}

\author[0000-0002-7464-498X]{E.~Zapartas}
\affiliation{IAASARS, National Observatory of Athens, Vas. Pavlou and
  I. Metaxa, Penteli, 15236, Greece}

\author[0000-0001-7969-1569]{S.~Justham}
\affiliation{School of Astronomy and Space Science, University of the
  Chinese Academy of Sciences, Beijing 100012, PR China}
\affiliation{Anton Pannekoek Institute of Astronomy and GRAPPA,
  University of Amsterdam, Science Park 904, 1098 XH Amsterdam, The
  Netherlands}
\affiliation{Max-Planck-Institut für Astrophysik, Karl-Schwarzschild-Straße 1, 85741 Garching, Germany}

\author[0000-0001-5228-6598]{K.~Breivik}
\affiliation{Center for Computational Astrophysics, Flatiron
  Institute, New York, NY 10010, USA}

\author[0000-0002-6592-2036]{M. Lau}
\affiliation{School of Physics and Astronomy, Monash University,
  Clayton, Victoria 3800, Australia}
\affiliation{OzGrav: The ARC Centre of Excellence for Gravitational Wave Discovery, Australia}
\affiliation{Center for Computational Astrophysics, Flatiron  Institute, New York, NY 10010, USA}

\author[0000-0003-3441-7624]{R.~Farmer}
\affiliation{Max-Planck-Institut für Astrophysik, Karl-Schwarzschild-Straße 1, 85741 Garching, Germany}

\author[0000-0002-8171-8596]{M.~Cantiello}
\affiliation{Center for Computational Astrophysics, Flatiron
  Institute, New York, NY 10010, USA}

\author[0000-0002-4670-7509]{B.~D.~Metzger}
\affiliation{Columbia Astrophysics Laboratory, Columbia University, New York, New York 10027, USA}
\affiliation{Center for Computational Astrophysics, Flatiron Institute, New York, NY 10010, USA}

\begin{abstract}
  Common-envelope (CE) evolution is an outstanding open problem in
  stellar evolution, critical to the formation of compact binaries
  including gravitational-wave sources. In the ``classical''
  isolated binary evolution scenario for double compact objects, the
  CE is usually the second mass transfer phase. Thus, the donor star
  of the CE is the product of a previous binary interaction, often
  stable Roche-lobe overflow (RLOF). Because of the accretion of mass
  during the first RLOF, the main-sequence core of the accretor star
  grows and is ``rejuvenated''. This modifies the core-envelope
  boundary region and decreases significantly the envelope binding
  energy for the remaining evolution. Comparing accretor stars from
  self-consistent binary models to stars evolved as single, we
  demonstrate that the rejuvenation can lower the energy required to
  eject a CE by $\sim 42-96\%$ for both black hole and neutron star
  progenitors, depending on the evolutionary stage and final orbital
  separation. Therefore, binaries experiencing first stable mass
  transfer may more easily survive subsequent CE events and
  result in possibly wider final separations compared to current predictions.
  Despite their high mass, our accretors also experience extended
  ``blue loops'', which may have observational consequences for
  low-metallicity stellar populations and asteroseismology.
\end{abstract}


\section{Introduction}
\label{sec:intro}

Common envelope (CE) evolution is important for massive
isolated binaries to become gravitational-wave (GW) sources, despite
recent debates on its relevance for the progenitors of the most
massive binary black holes \citep[e.g.,][]{vandenheuvel:2017,
  pavlovskii:2017, klencki:2020, klencki:2021, vanson:2021,
  marchant:2021}. CE remains a crucial step in the formation, among
many other compact binaries, of cataclysmic variable
\citep[e.g.,][]{paczynski:1976}, double white dwarfs
\citep[e.g.,][]{zorotovic:2010, korol:2017, kremer:2017, renzo:21gwce,
  thiele:21}, binary neutron stars \citep[NS,
e.g.,][]{vigna-gomez:2018, vigna-gomez:2020}, merging black
hole-neutron stars \citep[e.g.,][]{kruckow:18, broekgaarden:21}, and
possibly low-mass binary black holes \citep[BH, e.g.,][]{dominik:2012,
  vanson:2021}.

In the ``classical scenario'' for binary BHs and/or NSs
\citep[e.g.,][]{tutukov:93,belczynski:2016, tauris:2017}, the
progenitor binary experiences a first dynamically stable mass transfer
through Roche-lobe overflow (RLOF) between two non-compact stars.
Subsequently, the initially more massive RLOF-donor collapses to a
compact object without disrupting the binary
\citep[e.g.,][]{blaauw:1961,renzo:2019walk}. Only afterwards, as the
initially less massive RLOF-accretor expands, a second mass-transfer
phase occurs and it can be dynamically unstable, that is a CE
\citep[e.g.,][]{dominik:2012, belczynski:2016, kruckow:18}. This
second mass transfer is responsible for the orbital shrinking
\citep{paczynski:1976} allowing the system to merge within the age of
the Universe. Therefore, in this scenario, the donor star of the CE is
the former accretor of the first RLOF \citep[e.g.,][]{klencki:2020,
  law-smith:2020, renzo:2021zoph}.

The first stable RLOF typically occurs during the main sequence of the
initially less massive star and accretion modifies its structure
\citep[e.g.,][]{neo:1977, packet:1981, blaauw:1993, cantiello:2007,
  renzo:2021zoph}. On top of the enrichment of the envelope with
CNO-processed material from the donor star core \citep{blaauw:1993,
  renzo:2021zoph, el-badry:2022a}, and the substantial spin-up \citep[e.g.,][]{packet:1981},
accretors are expected to adjust their core-size to the new mass in a
``rejuvenation'' process \citep[e.g.,][]{neo:1977, hellings:1983,
  hellings:1984}. The readjustment is driven by mixing at the boundary
between the convective core and the envelope, which refuels the
burning region of hydrogen (H), increasing the stellar lifetime. This
mixing also affects the thermal structure of the partially H-depleted
layer above the helium-rich core (He), which we refer to as
core-envelope boundary (CEB) region. It is in the CEB that the density
rises and most of the envelope binding energy is accumulated for the
remaining stellar lifetime \citep[e.g.,][]{tauris:01, ivanova:2013,
  ivanova:2020}. Consequently, the success or failure of the CE
ejection, and the final separation, are likely decided in the CEB
layer and may be different depending on whether the CE-donor accreted
mass previously or not.

Here, we use structure and evolution binary models to study the impact
of the first RLOF phase on the outcome of possible subsequent CE
events. \Secref{sec:methods} describe our \mesa\ calculations. In
\Secref{sec:bin_models} we show the ratio of binding energy of our
accretor models divided by the binding energy of single stars with same
total post-RLOF mass. We discuss our findings and conclude in
\Secref{sec:conclusions}. Appendix~\ref{sec:toy_models} presents a
proof-of-principle numerical experiment illustrating the effect of
changing the CEB region and rotation on the envelope binding energy,
and Appendixes~\ref{sec:BE}--\ref{sec:pop_synth_app} present
additional plots of our model grids.

\section{Pre-common envelope evolution}
\label{sec:methods}

We use \mesa\ \citep[version 15140,][]{paxton:2011, paxton:2013,
  paxton:2015, paxton:2018, paxton:2019, jermyn:2022} to compute the evolution of
binaries which experience mass transfer after the end of the donor's
main sequence, that is case B RLOF
\citep[][]{kippenhahn:1967}. Our output files are compatible for
use in the population synthesis code \code{POSYDON}
\citep{fragos:2022} and publicly available together with our input
files and customized routines at
\href{https://zenodo.org/record/7036016}{doi:10.5281/zenodo.7036016}.
Our setup is similar to \cite{renzo:2021zoph}, except for the
metallicity: here we adopt $Z=0.0019\simeq Z_\odot/10$, relevant for
the progenitor population of GW events \citep[e.g.,][]{vanson:2021}.
Moreover, we apply throughout the star a small amount of mixing with
diffusivity \texttt{min\_D\_mix}=0.01\,$\mathrm{cm^2 s^{-1}}$. This
improves the numerical stability by smoothing properties across
adjacent cells, without introducing significant quantitative
variations, and is a typical numerical technique used in
asteroseismology calculations (J.~Fuller, private~comm.).

We adopt an initial period $P=100$\,days and choose initial masses
$(M_{1}, M_{2}) = (18, 15), (20, 17), (38, 30)\,M_\odot$. We focus on
the initially less massive stars, which after accretion become
$M_2=15\rightarrow 18, 17\rightarrow 20, 30\rightarrow 36\,M_\odot$,
roughly representative of NS, uncertain core-collapse outcome, and BH
progenitors, respectively. However, the core-collapse outcome (NS or
BH formation, with explosion or not), cannot be decided solely based
on the (total or core) mass of a star \citep[e.g.,][]{oconnor:11,
  farmer:16, patton:2020, zapartas:21b, patton:22}.

During the binary evolution, we account for tidal
  interactions assuming each stellar layer reacts on its own timescale
  (see \citealt{paxton:2015}). At mass transfer, our \mesa\ models
assume that the accretion efficiency is limited by
rotationally-enhanced wind mass loss \citep[e.g.,][]{sravan:2019,
  wang:2020, renzo:2021zoph, sen:2022}. However, this may lead to less
conservative mass transfer than suggested by observations
\citep[e.g.,][]{wang:2021a}.

After the donor detaches from the Roche lobe, our simulations
artificially separate\footnote{We make the routine to separate a
  \mesa\ binary on-the-fly publicly available
  \url{https://github.com/MESAHub/mesa-contrib/}} the stars and
continue the evolution of the accretor as a single star until it
reaches carbon depletion (defined by central carbon mass fraction
$X_\mathrm{c}(^{12}\mathrm{C})<2\times10^{-4} $). ducing
  the complexity by not simulating the late evolutionary phases of the
  RLOF-donors saves significant computing time at a small price in
  accuracy of the RLOF-accretors. Separating the stars, we
  neglect further possible, but not expected, mass-transfer episodes
  (case BB RLOF, \citealt{delgado:81, laplace:2020}). We also neglect
  post-RLOF tides, which are expected to be
  negligible for wide pre-CE binaries. Finally, we ignore the impact of
  the donor's supernova ejecta with the accretor \cite[which has a small and
  short-lasting effect only on the outermost layers,
  e.g.,][Hirai~R., private communication]{hirai:2018, ogata:2021} and the orbital consequences of the
core-collapse \citep[e.g.,][]{brandt:1995, kalogera:1996,
    tauris:1998, renzo:2019walk}. To illustrate the physical reason
why the first RLOF may influence the envelope structure of the
accretor much later on, we also compute comparison stars. For each
mass, we compute non-rotating single stars with otherwise identical
setup, and ``engineered'' stars which we modify at terminal age main
sequence (TAMS, central hydrogen mass fraction
$X_\mathrm{c}(^1\mathrm{H})<10^{-4}$) to mimic crudely the impact of
rejuvenation of the accretors CEB (see Appendix~\ref{sec:toy_models}).

At the onset of a CE event, the photospheric radius
$R\equiv R_\mathrm{RL, donor}$ is the size of the Roche lobe of the
donor star determined by the binary separation and mass ratio
\citep[e.g.,][]{paczynski:1971, eggleton:83}. Thus, we compare the
internal structure of accretors to single and engineered stars at
various epochs defined by a fixed photospheric radius
$R=100,\ 200,\ 300,\ 500,\ 1000\,R_\odot$.

\section{Accretors from self-consistent binary models}
\label{sec:bin_models}

\begin{figure}[tbp]
  \includegraphics[width=0.5\textwidth]{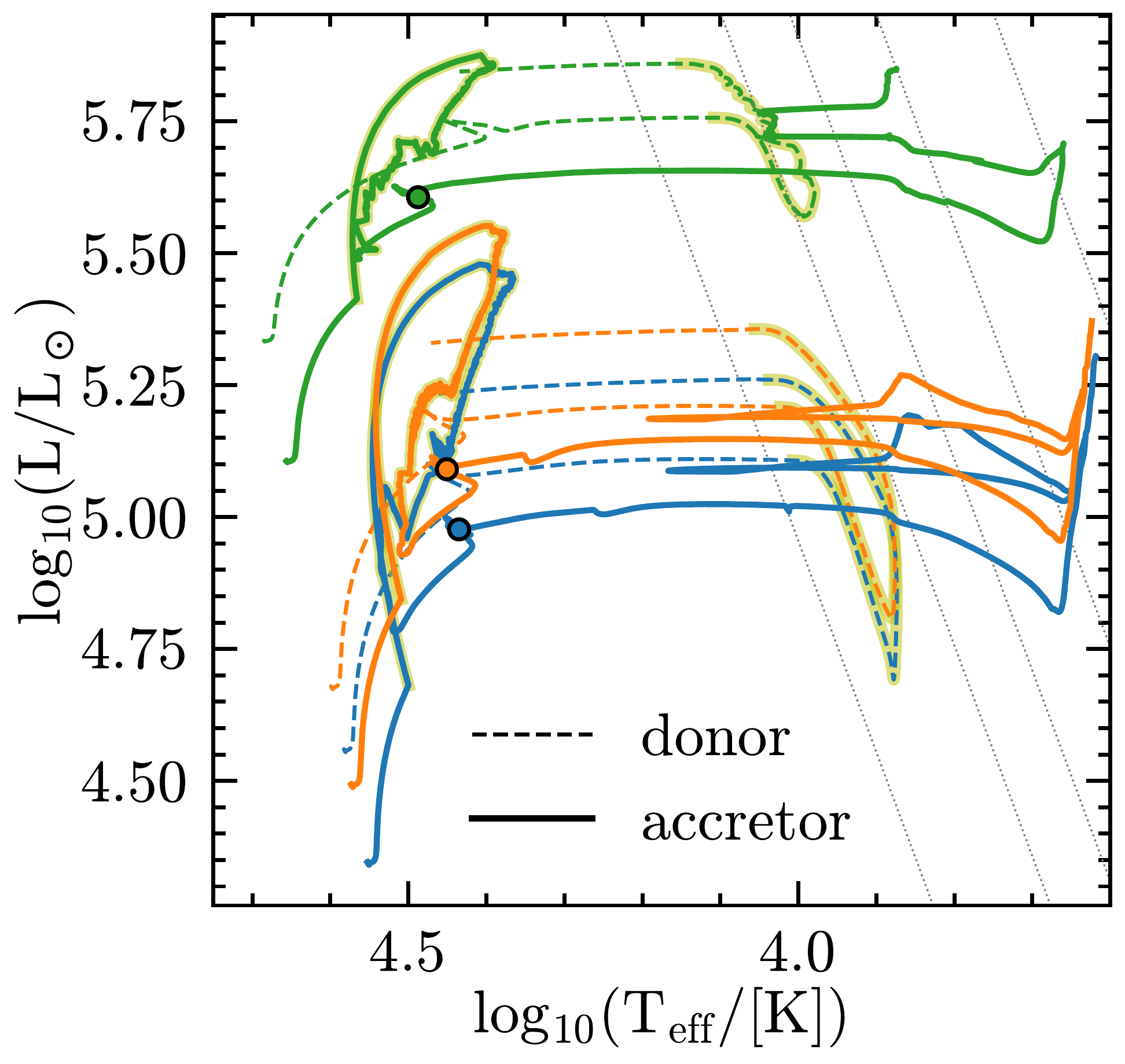}
  \caption{HR diagram of the binary systems. The thin dashed lines
    show the evolution of the donors until RLOF detachment, the solid
    lines show the accretors from ZAMS, through RLOF (marked by a
    yellow outline), until core carbon depletion. Dots with black
    outlines mark the accretor's TAMS (not shown for donor). The thin
    dotted lines mark constant radii of
    $R=100, 200, 300, 500, 1000\,R_\odot$, all models have $Z=0.0019$,
    an initial orbital period of $100$\,days, and initial masses of 38
    and 30\,$M_\odot$ (green), 20 and 17\,$M_\odot$ (orange), 18 and
    15\,$M_\odot$ (blue).}
  \label{fig:HRD}
  \script{HRD.py}
\end{figure}

\Figref{fig:HRD} shows the evolution of our binaries on the
Hertzsprung-Russell (HR) diagram. The thin dashed lines show the
evolution of the donor stars \citep[e.g.,][]{morton:60, gotberg:2018,
  laplace:2021} from zero age main sequence (ZAMS), through RLOF,
until our definition of detachment. The solid lines correspond to the
full evolution of the accretors, from ZAMS, through RLOF, until carbon
depletion. The yellow outline of the tracks highlight the RLOF mass
transfer (see e.g., \citealt{renzo:2021zoph}). During this phase the
accretor progressively spins-up, and accretes CNO-processed material
from the donor's inner layers which are mixed downwards in the
envelope by meridional circulations and thermohaline mixing, and its
core is rejuvenated because of the increased mass \citep[see
also][]{sravan:2019, renzo:2021zoph, wang:2020}. During the brief RLOF
phase, our accretors grow to
$M_2=15\rightarrow 18, 17\rightarrow 20, 30\rightarrow 36\,M_\odot$,
respectively, corresponding to an overall mass transfer efficiency
$\beta_\mathrm{RLOF}=|\Delta M_\mathrm{accretor}/\Delta M_\mathrm{donor}| = 0.29,\ 0.30,$\,and
0.43, respectively \citep[see discussion in ][]{renzo:2021zoph}.
The binaries started with an initial separation of
  $\sim{}300\,R_\odot$ and widen to $\sim{}380\,R_\odot$ days by RLOF
  detachment. We expect further widening caused by the wind mass loss
  of both stars, allowing us to neglect tides in the remaining
  evolution and the impact of the RLOF-donor collapse
  \citep{hirai:2018, ogata:2021}.

All three accretor models experience a blueward evolution after
beginning to ascend the Hayashi track. In the two lowest mass models,
this results in a blue-loop, which last $\sim{}10^5$ years. These
models spend a significant fraction of their He core burning as hot
yellow/blue supergiants, and reach
$\log_{10}(T_\mathrm{eff}/\mathrm{[K]})\gtrsim 4.2$. Our most massive
accretor ($M_2=30\rightarrow 36\,M_\odot$) evolves towards hotter
temperatures during core He burning, but never fully recovers closing
the blue loop. Its excursion to hottest temperatures occurs after He
core depletion and lasts $\sim{}10^{4}$ years.

Blue loops are not expected for single stars with
$M\gtrsim 12\,M_\odot$ \citep[e.g.,][]{walmswell:2015}, and their
occurrence is known to be sensitive to the He profile above the
H-burning shell, and specifically the mean molecular weight profile
\citep{walmswell:2015, farrell:22}. Thus it is not surprising that
RLOF-accretion, which modifies the CEB, may lead to blue loops, and
formation of yellow supergiants. We note that comparison single stars
also experience late blue-ward evolution, but not a ``loop'' back to
red. This behavior is likely related to the relatively high wind
mass-loss rate assumed (see \citealt{renzo:2017}), and the models with
initial mass $\gtrsim 30\,M_\odot$ are qualitatively similar to the
most massive accretor in \Figref{fig:HRD} even without accreting
matter from a companion: the occurrence of blue loops is notoriously
sensitive to many single-star physics uncertainties, and while they
appear consistently in our accretor models, their physicality should
be tested further.

However, in the context of CE progenitors, blue loops are not crucial
since they correspond to a decrease in radius, which would not result
in binary interactions during the loop. They might change the
mass-loss history of the accretor, but since they occur in a short
evolutionary phase, their impact should be
limited.

\begin{figure*}[htbp]
  \centering
  \includegraphics[width=\textwidth]{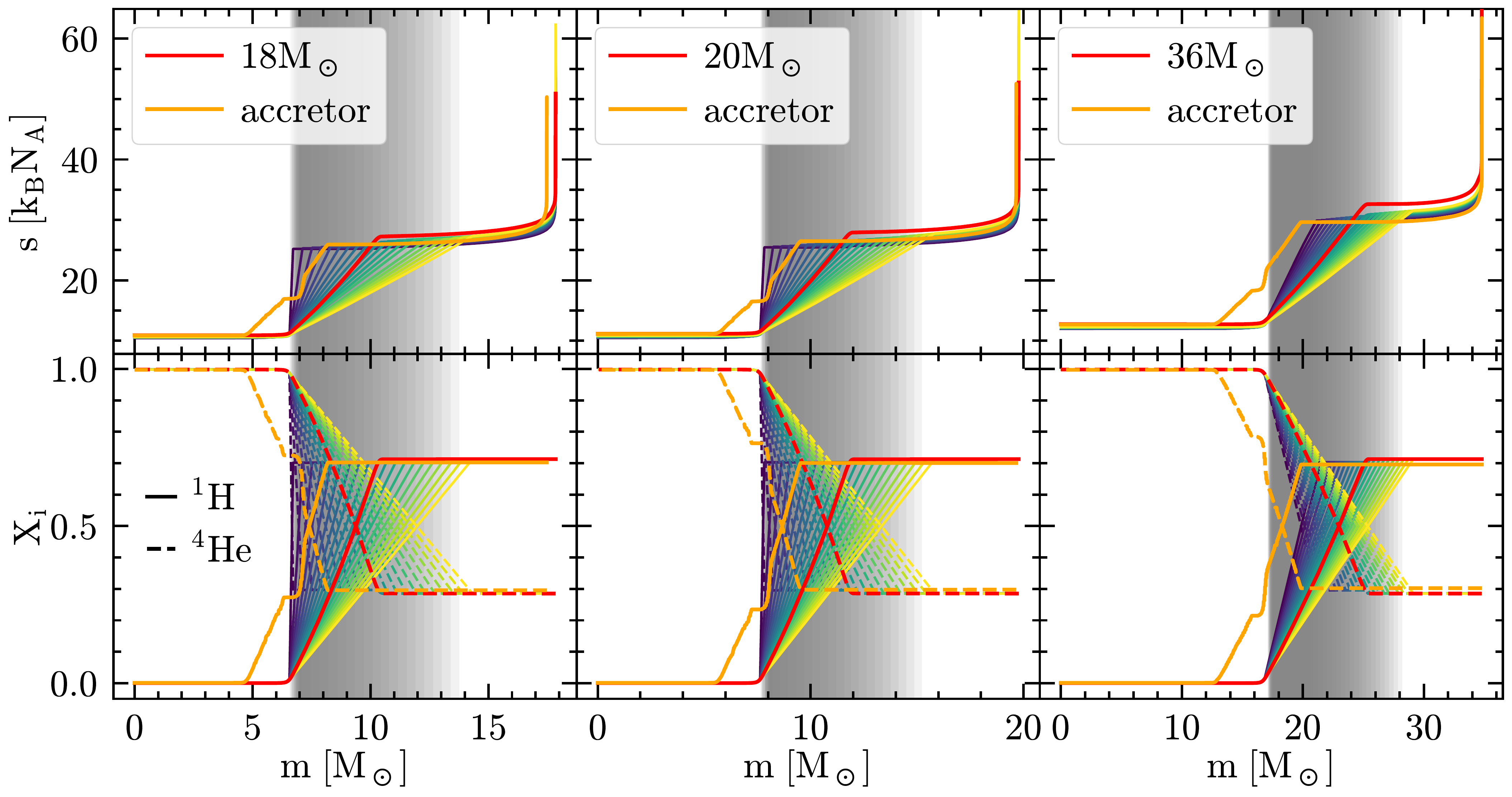}
  \caption{Specific entropy s (top row), H (bottom row, solid lines),
    and He (bottom row, dashed lines) TAMS profiles for non-rotating
    single stars (red), accretors (orange), and ``engineered'' models
    of the same total mass as the post-RLOF mass of the accretors. The
    overlapping gray bands emphasize the CEB region,
    $X_\mathrm{c}(^1\mathrm{H})+0.01<X(^1\mathrm{H})<X_\mathrm{surf}(^1\mathrm{H})-0.01$,
    with $X_c$ and $X_\mathrm{surf}$ the central and surface value of
    the hydrogen mass fraction. The CEB size of the engineered models
    increases from blue to yellow.}
  \label{fig:TAMS_profiles}
  \script{TAMS_profiles.py}
\end{figure*}

\Figref{fig:TAMS_profiles} shows the specific entropy (s) profile --
which determines the instantaneous dynamical response of the gas --
and the H and He mass fractions at TAMS for our accretor models
(orange), single non-rotating stars (red) and ``engineered'' models of
roughly same total mass as the accretor post-RLOF. A gray region
highlights the CEB, and their overlap produces the shade in
\Figref{fig:TAMS_profiles}. We compare our models at the same total
post-RLOF mass ($M\simeq M_{2}$) because it enters in \Eqref{eq:BE}
and is typically used in rapid population synthesis codes to construct
accretors from single star models \citep[e.g.,][]{hurley:2002,
  breivik:2020}. We
present in \Figref{fig:TAMS_profiles_same_initial_mass} a comparison
between TAMS profiles of accretors, single stars, and engineered
models with the same \emph{initial} mass as an alternative comparison
that should bracket the range of sensible comparison models.

\begin{figure*}[htbp]
  \includegraphics[width=\textwidth]{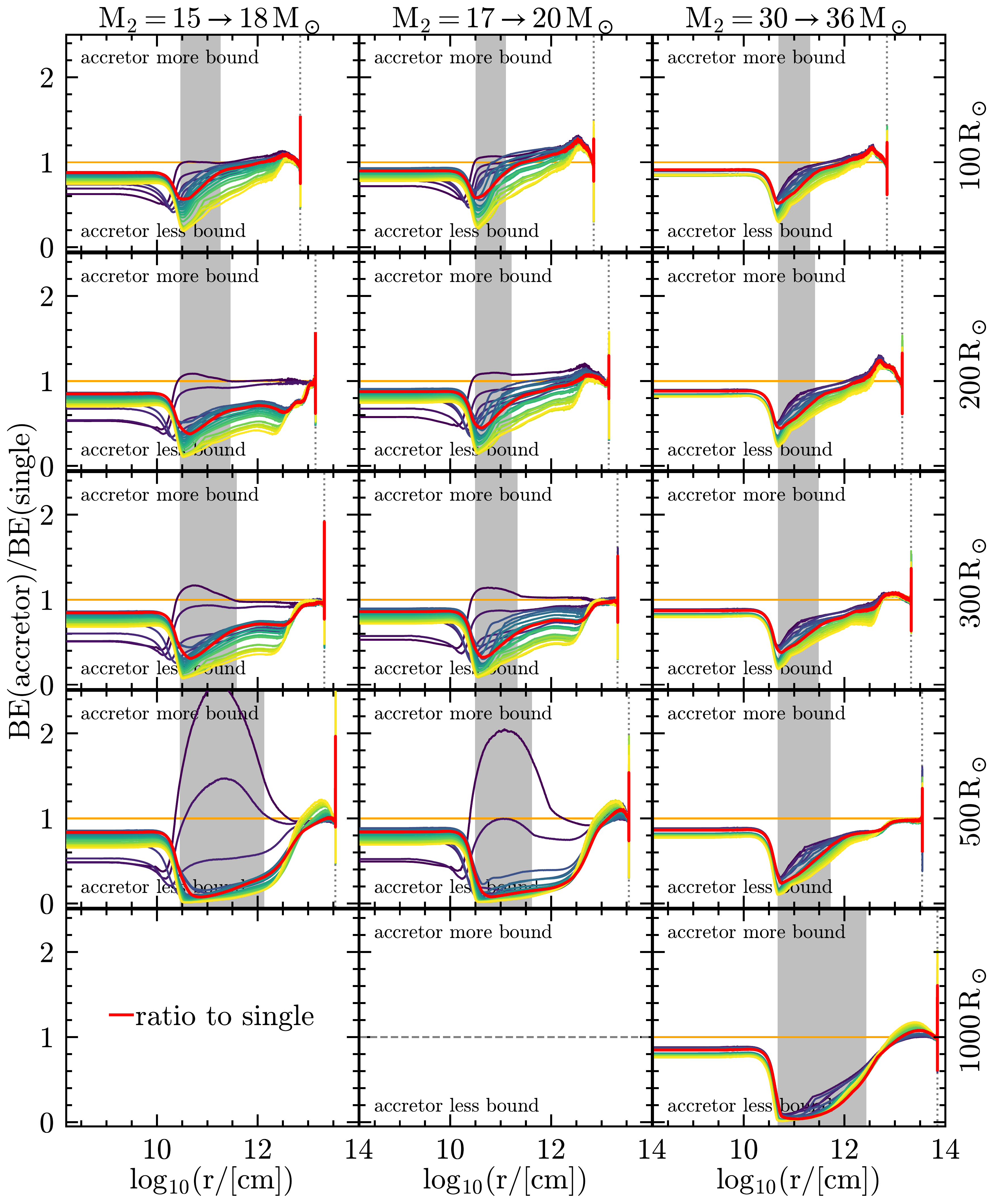}
  \caption{Ratios of the binding energy profiles (including internal
    energy, $\alpha_\mathrm{th}=1$) of the accretor stars divided the
    binding energy profile of stars of the same total mass post-RLOF.
    The orange solid line at 1 shows the ratio of the accretor to
    itself as a check on the models interpolation, red solid lines
    show the ratio of the accretor to a non-rotating single star,
    while the other colors show the ratio to ``engineered'' stars
    (see~\Figref{fig:TAMS_profiles}, increasing CEB
      size from blue to yellow, see
    also Appendix~\ref{sec:eng_examples}). Each panel shows the ratios
  at the first time the models reach the radius indicated on the right
  and by the vertical dotted gray lines. The vertical
  gray bands mark the radial extent of the CEB in the accretors only, which
  may differ in the other stars. For the binding energy
  profiles in the numerator and denominator of the fractions plotted
  here see \Figref{fig:BE_profiles}.}
  \label{fig:grid_ratios}
  \script{grid_ratios.py}
\end{figure*}

Because of the timing and duration of RLOF, accretion affects the CEB
layers in more subtle ways than we impose in our ``engineered''
models. One expects the CEB in accretors to be steeper than in a star
evolving as single, resulting in models qualitatively more similar to
our engineered models with the steeper entropy and composition in the
CEB (darker lines in \Figref{fig:TAMS_profiles}-\ref{fig:grid_ratios},
\Figref{fig:toy_models_example}, and
\Figref{fig:TAMS_profiles_same_initial_mass}-\ref{fig:lambda_grid}).
The convective core of the accretor post-RLOF would naturally become
more massive in a star with homogeneous composition. However, the
He-enriched CEB can impede or prevent the growth of the core
\citep[e.g.,][]{yoon:2005}. The He-enrichment increases with the
stellar age, and thus with the duration of the pre-RLOF evolution.
This duration depends on the binary architecture: for our binaries
with initial $P=100$\,days and $q=M_2/M_1\simeq 0.8$, RLOF starts
after $\sim{}$10, 9, and 5\,Myrs from the least massive to the most
massive system, which correspond to central H mass fractions
$X_\mathrm{c}(^1\mathrm{H}) = 0.27, 0.23$, and $0.21$ for the
accretors (cf.~height of the plateaus in the orange lines in
\Figref{fig:TAMS_profiles}). Our oversimplified engineered models do
not exhibit such plateau because they are constructed assuming
instantaneous rejuvenation at TAMS (see
Appendix~\ref{sec:eng_examples}).

To quantify the impact of the first RLOF phase on the outcome of the
second mass transfer phase, we evolve forward all the TAMS profiles
shown in \Figref{fig:TAMS_profiles} and compare them at fixed outer
radii. In the ``classical'' binary evolution path, after the
RLOF-donor collapses to a compact object, the evolutionary expansion
of the RLOF-accretor triggers a CE. This phase of evolution is a
complicated physics problem, not necessarily well-described as an
energetically closed system \citep[e.g.,][]{ivanova:2013,
  ivanova:2020, renzo:21gwce}.  However, a common oversimplification
is to assume energy conservation
\citep[``$\alpha_{\mathrm{CE}}\lambda_\mathrm{CE}$ algorithm'',
e.g.,][]{webbink:1984, dekool:1990, demarco:11} to determine CE
ejection and final separation. Here we focus on the
RLOF-accretor/CE-donor binding energy profile as an indication for the
ease of CE ejection. Even if imperfect, following
  common practice, we adopt this quantity as a
proxy for the physical processes which determine the CE outcome
and that allows us to compare models to each other. We
calculate the cumulative binding energy outside mass coordinate $m$ as
\citep[e.g.,][]{dekool:1990, dewi:2000, lau:2022}:
\begin{equation}
  \label{eq:BE}
BE(m, \alpha_{\rm th}) = - \int_{m}^M\,dm'\left( -\frac{G m'}{r(m')}+\alpha_\mathrm{th} u(m')\right) \ \ ,
\end{equation}
with $r(m')$ radius, $u(m')$ the internal energy of a shell of mass
thickness $dm'$ and outer Lagrangian mass coordinate $m'$, and $G$ the
gravitational constant. The integral goes from mass coordinate $m$,
which can be thought of as the mass of the ``core'' surviving a
hypothetical CE, to the surface. The parameter
$0\leq \alpha_\mathrm{th}\leq 1$ is the fraction of internal energy
(including recombination energy) that can be used to lift the shared
CE \citep[e.g.,][]{han:95}. It is possible that $\alpha_\mathrm{th}$
may not be constant during a CE (e.g., if recombination happens in
already unbound material it cannot contribute to the CE energetics,
\citealt{lau:22}) or across binary systems entering a CE at different
evolutionary stages. For $\alpha_\mathrm{th}=0$, \Eqref{eq:BE} give
the gravitational binding energy (dashed lines in
\Figref{fig:toy_models_example}-\ref{fig:rotation_models_example}),
while $\alpha_{\mathrm{th}}=1$ assumes perfectly fine-tuned use of all
the internal energy \citep[solid lines, see also][]{klencki:2020}. These two
cases bracket the range of possible use of internal energy to eject
the CE. The additional inclusion of a rotational-energy term
$0.5 \mathcal{I} \omega^2$ (with $\mathcal{I}=2r^2/3$ the specific
moment of inertia) in the integral in \Eqref{eq:BE} contributes to
less than $\lesssim 10\%$ of the cumulative binding energy only in the
outermost layers -- likely to be crossed during a dynamical plunge-in
in CE evolution -- and only for $R\lesssim 300\,R_\odot$ --
afterwards, even the accretor spin down significantly (see also
Appendix~\ref{sec:rot_examples}).

Because of the large range of $BE$ across the stellar structures, it
is hard to appreciate directly the magnitude of the effect of RLOF-driven
rejuvenation on the $BE$ profile (shown in \Figref{fig:BE_profiles}). \Figref{fig:grid_ratios} presents the
ratio of the local value of the cumulative binding energy from the
surface of our accretor models divided by the
comparison single stars, as a function of radius. The two
lowest mass accretors (left and central column) do not expand to
$R=1000\, R_\odot$ before carbon depletion. To compute the
ratio, we interpolate linearly the single star models on the mesh of
our accretor, using the fractional Lagrangian mass coordinate $m/M$ as
independent coordinate.
We calculate these ratios
when both the stars reach for the first time radii
$R=100, 200, 300, 500, 1000\,R_\odot$ (see vertical gray dotted
lines), corresponding to the assumed Roche lobe radius of the donor at
the onset of the CE.

In each panel, radial coordinates $r$ for which the lines in
\Figref{fig:grid_ratios} are below one correspond to radii at which
the accretor models are less bound than the comparison single star or
engineered model. For $R\lesssim 300R_\odot$, the outermost layers
(more likely to be crossed by the binary during the dynamical
plunge-in phase of the CE) may be slightly less bound in single stars
than accretors (red line greater than 1) -- partly because of
the impact of rotation. But for most of the envelope
radius, the ratio is smaller than one, suggesting it would take less
energy to eject the outer layers of the envelope of the accretors down
to such $r$. All of our accretor models, regardless of them being NS
or BH progenitor, and regardless of their evolutionary phase,
are qualitatively more similar to the darker lines
representing engineered models with steeper CEB profiles.

\begin{figure}[htbp]
  \centering
  \includegraphics[width=0.5\textwidth]{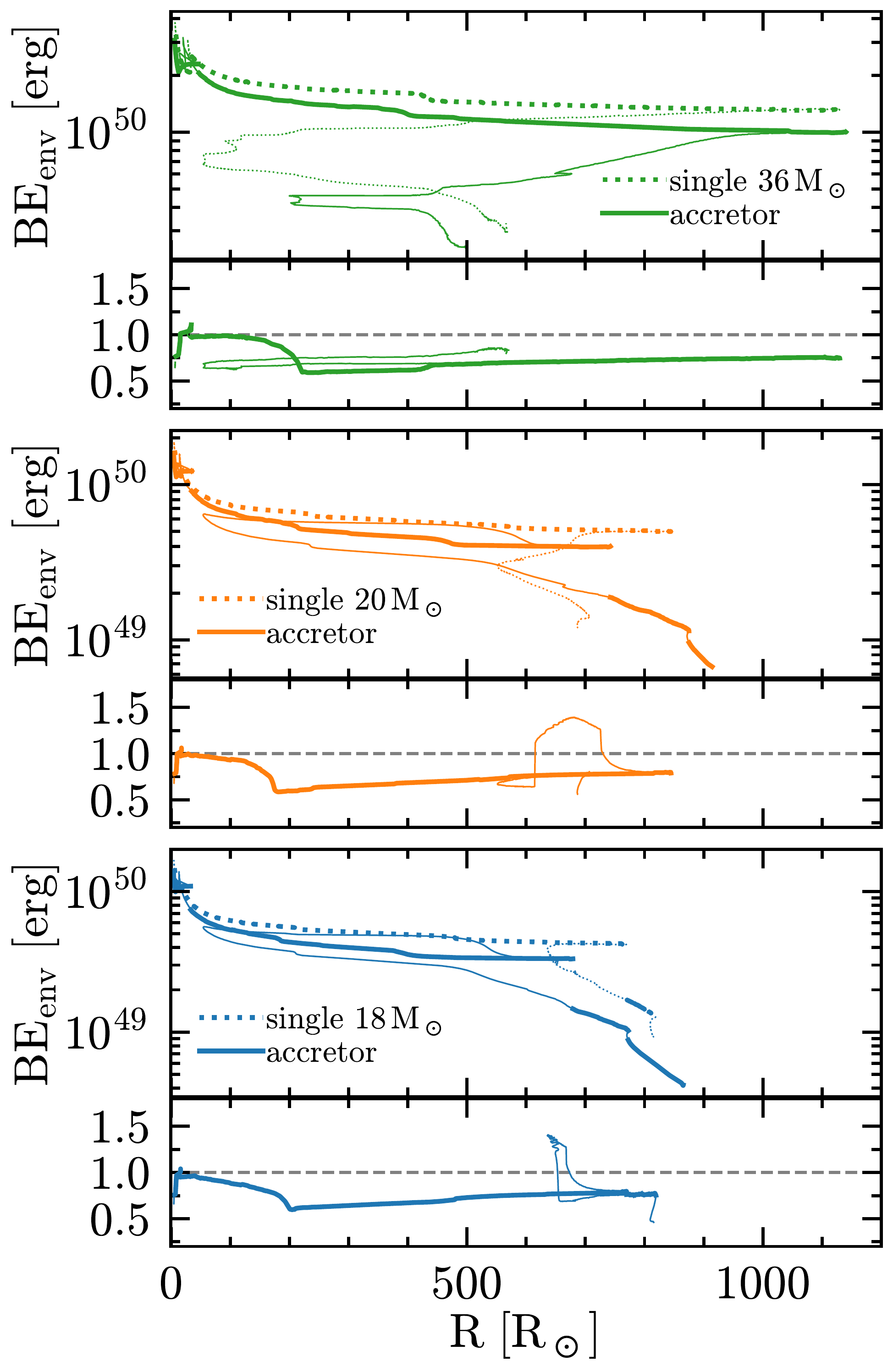}
  \caption{Evolution as a function of the photospheric radius $R$ of
    the binding energy (including the thermal energy,
    $\alpha_\mathrm{th}=1$) of the accretors and single stars of the
    same (post-RLOF) total mass. The solid and dotted lines show the
    accretors and single stars, respectively, and they are thin when
    previously the models reached larger radii. The bottom panels show
    the ratio of the binding energies, which is always smaller than 1
    the first time a certain radius $R$ is reached (thick lines),
    indicating that the accretors might have envelopes easier to
    unbind in at the start of a CE event. Excursions above 1 of the
    ratio occur only during phases of radius decrease (thin lines)
    which could not initiate a CE. See \Figref{fig:BE_profiles} for
    the radial binding energy profile at given outer radii $R$.}
  \label{fig:BE_env_R}
  \script{BE_env_R.py}
\end{figure}

The minimum ratio of binding energies occur roughly at the inner edge
of the CEB layer in \Figref{fig:grid_ratios}. Considering the
ratio to single stars (red lines), the minima range between 0.56-0.07,
0.58-0.08, and 0.51-0.04 from our least to most massive binary. In
other words, at the radius where the difference between accretors and
single stars models is largest, which is also the location where the
outcome of a common envelope is likely to be decided, the accretor's
binding energy is roughly between $\sim{}50-\mathrm{few}\%$ of the
binding energy of a single star. Regardless of the mass, the larger
the outer radius the smaller the minimum of the ratio of binding
energies: the differences caused by RLOF accretion and rejuvenation of
the core grows as stars evolve and their core contracts.

Defining the He core boundary as the outermost location where $X<0.01$
and $Y>0.1$, we can fix $m=M_\mathrm{He}$ in \Eqref{eq:BE} to obtain
an integrated binding energy for the envelope:
\begin{equation}
  \label{eq:BE_env}
  BE_\mathrm{env} \equiv BE(m=M_\mathrm{He}, \alpha_{th}=1) \ \ .
\end{equation}
\Figref{fig:BE_env_R} shows the evolution of this integrated envelope
binding energy as a function of the outer radius. Each panel shows one
of our binaries, from top to bottom: 36+30\,$M_\odot$,
20+17\,$M_\odot$, 18+15\,$M_\odot$. For each binary, the lower panel
shows the ratios of the envelope binding energy of the accretor
divided by the binding energy of the comparison single star (i.e., the
ratio of the solid lines to the dotted lines in the
panel above). To compute these ratios, we interpolate our accretor
models on the time-grid of the single stars using the central
temperature $\log_{10}(T_c/[\mathrm{K}])$ as independent coordinate.
In each of the lower panels the ratios are lower than one (marked by
the gray dashed lines) suggesting that post-RLOF accretor stars have
envelopes that require less energy to be ejected in a CE event. The
only times the binding energy of the accretor is higher than the
corresponding single stars is during the blue-ward evolution discussed
earlier, which would not trigger a CE.

\section{Discussion \& Conclusions}
\label{sec:conclusions}

We have modeled the impact of mass transfer on the envelope structure
of the accretor, focusing on thermal timescale, post-donor-main
sequence case B RLOF (see \Figref{fig:HRD}). The accretion of mass
drives the growth of the accretor's core, changing the core/envelope
boundary region and ``rejuvenating'' the star (\Figref{fig:TAMS_profiles}). As the accretors evolve
beyond the main sequence, they experience large blue-loops which are
not expected in single stars of the same mass -- with potential
implications for asteroseismology \citep[e.g.,][]{dorn-wallenstein:20},
and the search for non-interacting companions to compact objects
\citep[e.g.,][]{breivik:17, andrews:19, chawla:21}.

% stellar hydro
The rejuvenation is driven by convective core boundary mixing
\citep[e.g.,][]{hellings:1983, hellings:1984, cantiello:2007,
  renzo:2021zoph}, and does not occur in its absence \citep{braun:95}.
The hydrodynamics of convective boundaries in stellar regime is an
active topic of research \citep[e.g.,][]{anders:22a, anders:22b}, and
observations of the width of the main sequence
\citep[e.g.,][]{brott:11} and asteroseismology
\citep[e.g.,][]{moravveji:16} suggest the presence of convective
boundary mixing in the core of massive main-sequence stars. In our
one-dimensional accretor models, the dominant core boundary mixing is
overshooting, with rotationally driven instabilities contributing to a
lesser extent during late RLOF. We adopt an exponentially decreasing
overshooting diffusion coefficient (\citealt{claret:17}) which may
underestimate the amount of mixing at the accretor core boundary.
After RLOF, a thick convective shell develops above the core
\citep[see][]{renzo:2021zoph}, which also contributes to the different
binding energy profiles (see \Figref{fig:BE_profiles}).

% CE other uncertainties
We have focused on the structural consequences of RLOF accretion,
specifically their impact on the subsequent binary interaction in the
``classical'' scenario to a GW merger: the CE event initiated by the
RLOF-accretor. Accretors have overall lower binding energy of the envelope
(both integrated from the surface to the He core,
see~\Figref{fig:BE_env_R} and as a function of radius,
see~\Figref{fig:grid_ratios}, \ref{fig:BE_profiles}, and \ref{fig:lambda_grid}). The
systematically lower binding energy our accretor models compared to
single stars of the same outer radius and total mass may
imply easier to eject (post-RLOF, second) CE and wider post-CE separations.

Before the onset of the dynamical instability in a CE event, a pre-CE
thermal timescale phase of mass transfer may occur
\citep[e.g.,][]{hjellming:1987, nandez:14, pejcha:17,
  blagorodnova:2021}. This phase may impact the envelope structure of
CE-donors (through tidal interactions and mass loss) whether they are
RLOF-accretors (as in our models) or not. Multidimensional studies are
needed to assess whether rejuvenation, rotation, tides, and the impact
of the companion's supernova shock on accretor stars counteract or
compound each other.

A key uncertainty in CE outcome is the location of the separation
between the (possibly) ejected envelope and the remaining core
\citep[e.g.,][]{tauris:01}. This affects equally each CE donor and
can have an amplitude comparable to the effect of rejuvenation in
accretors (see \Figref{fig:BE_profiles} for three possible definition
of ``core''). In the case of rejuvenated CE donors, uncertainties in the
core definition compound with the effect of rejuvenation itself.

% other binary architectures
Not all binary architectures necessarily result in rejuvenated
accretors like the ones described here. Very massive BH progenitors
($M_{\rm ZAMS}\gtrsim 40\,M_\odot$) may not expand as red supergiants
at all or avoid unstable mass transfer \citep[e.g.,][]{vanson:2021,
  marchant:2021}. Since their main-sequence lifetimes are roughly
independent of mass ($\sim{}2.5-3$\,Myr), at the first RLOF, accretors
this massive may already have a deep core/envelope chemical gradient
to prevent rejuvenation. However, more massive stars are generally
easier to mix \citep[including reaching rotationally-induced
chemically homogeneous evolution, e.g.,][]{yoon:2005, demink:2016}.
Shorter initial periods (i.e., earlier mass transfer)
and smaller radii during the binary interactions can
  prevent the red supergiant phase even at lower masses
\citep{cantiello:2007}.

We have focused on accretor models for progenitors of NS and BH. However,
the physical processes described should be similar in all accretor
stars with convective main sequence cores, down to initial mass
$M_{\rm ZAMS}\gtrsim 1.2\,M_\odot$ \citep[see also][]{wang:2020}. Thus,
also a fraction of progenitors of binaries with white dwarfs, if
sufficiently massive and experiencing a (case B) RLOF phase of
evolution, may be influenced by the structural differences between
single stars and RLOF-accretors.

%pop synth application
Including the structural reaction to accretion during RLOF in
population synthesis simulations could impact the distribution of
post-CE orbital separations, the predicted number of ``reverse''
stellar mergers \citep[e.g.,][]{zapartas:2017}, and the rate of GW
mergers. Our grid consists only of three binaries, but could be
extended to inform semi-analytic approximations of the binding energy
of CE-donor that have accreted mass in a previous stable mass transfer
phase (see also \Figref{fig:lambda_grid}).

\software{MESA \citep{paxton:2011, paxton:2015, paxton:2018,
    paxton:2019}, pyMESA \citep{pymesa}, \code{POSYDON} \citep{fragos:2022},
  compare\_workdir\_MESA\footnote{\url{https://github.com/mathren/compare_workdir_MESA/releases/tag/2.0}},
  Ipython \citep{ipython}, numpy
  \citep{numpy}, scipy \citep{scipy}, matplotlib \citep{matplotlib}, \showyourwork
  \citep{showyourwork}.}

% \vspace*{-40pt}
\acknowledgements{We are grateful to Y.~G\"otberg for
  invaluable discussions on binary physics, R.~Luger for the
  development of \showyourwork, and to the \code{POSYDON} team for
  sharing their \code{MESA} setup. EZ acknowledges funding support
  from the European Research Council (ERC) under the European Union’s
  Horizon 2020 research and innovation programme (Grant agreement No.
  772086). SJ acknowledges funding via the NWO Vidi research program BinWaves (project 639.042.728, PI: de Mink).}

\bibliographystyle{aasjournal}
\bibliography{./CE_accretors.bib}

\begin{thebibliography}{}
\expandafter\ifx\csname natexlab\endcsname\relax\def\natexlab#1{#1}\fi
\providecommand{\url}[1]{\href{#1}{#1}}
\providecommand{\dodoi}[1]{doi:~\href{http://doi.org/#1}{\nolinkurl{#1}}}
\providecommand{\doeprint}[1]{\href{http://ascl.net/#1}{\nolinkurl{http://ascl.net/#1}}}
\providecommand{\doarXiv}[1]{\href{https://arxiv.org/abs/#1}{\nolinkurl{https://arxiv.org/abs/#1}}}

\bibitem[{{Anders} {et~al.}(2022{\natexlab{a}}){Anders}, {Jermyn}, {Lecoanet},
  \& {Brown}}]{anders:22a}
{Anders}, E.~H., {Jermyn}, A.~S., {Lecoanet}, D., \& {Brown}, B.~P.
  2022{\natexlab{a}}, \apj, 926, 169, \dodoi{10.3847/1538-4357/ac408d}

\bibitem[{{Anders} {et~al.}(2022{\natexlab{b}}){Anders}, {Jermyn}, {Lecoanet},
  {Fraser}, {Cresswell}, {Joyce}, \& {Fuentes}}]{anders:22b}
{Anders}, E.~H., {Jermyn}, A.~S., {Lecoanet}, D., {et~al.} 2022{\natexlab{b}},
  \apjl, 928, L10, \dodoi{10.3847/2041-8213/ac5cb5}

\bibitem[{{Andrews} {et~al.}(2019){Andrews}, {Breivik}, \&
  {Chatterjee}}]{andrews:19}
{Andrews}, J.~J., {Breivik}, K., \& {Chatterjee}, S. 2019, \apj, 886, 68,
  \dodoi{10.3847/1538-4357/ab441f}

\bibitem[{Belczynski {et~al.}(2016)Belczynski, Holz, Bulik, \&
  O'shaughnessy}]{belczynski:2016}
Belczynski, K., Holz, D.~E., Bulik, T., \& O'shaughnessy, R. 2016, Nature, 534,
  512, \dodoi{10.1038/nature18322}

\bibitem[{{Blaauw}(1961)}]{blaauw:1961}
{Blaauw}, A. 1961, \bain, 15, 265

\bibitem[{{Blaauw}(1993)}]{blaauw:1993}
{Blaauw}, A. 1993, in Astronomical Society of the Pacific Conference Series,
  Vol.~35, Massive Stars: Their Lives in the Interstellar Medium, ed. J.~P.
  {Cassinelli} \& E.~B. {Churchwell}, 207

\bibitem[{{Blagorodnova} {et~al.}(2021){Blagorodnova}, {Klencki}, {Pejcha},
  {Vreeswijk}, {Bond}, {Burdge}, {De}, {Fremling}, {Gehrz}, {Jencson},
  {Kasliwal}, {Kupfer}, {Lau}, {Masci}, \& {Rich}}]{blagorodnova:2021}
{Blagorodnova}, N., {Klencki}, J., {Pejcha}, O., {et~al.} 2021, \aap, 653,
  A134, \dodoi{10.1051/0004-6361/202140525}

\bibitem[{Brandt \& Podsiadlowski(1995)}]{brandt:1995}
Brandt, N., \& Podsiadlowski, P. 1995, \mnras, 274, 461,
  \dodoi{10.1093/mnras/274.2.461}

\bibitem[{{Braun} \& {Langer}(1995)}]{braun:95}
{Braun}, H., \& {Langer}, N. 1995, \aap, 297, 483

\bibitem[{{Breivik} {et~al.}(2017){Breivik}, {Chatterjee}, \&
  {Larson}}]{breivik:17}
{Breivik}, K., {Chatterjee}, S., \& {Larson}, S.~L. 2017, \apjl, 850, L13,
  \dodoi{10.3847/2041-8213/aa97d5}

\bibitem[{{Breivik} {et~al.}(2020){Breivik}, {Coughlin}, {Zevin}, {Rodriguez},
  {Kremer}, {Ye}, {Andrews}, {Kurkowski}, {Digman}, {Larson}, \&
  {Rasio}}]{breivik:2020}
{Breivik}, K., {Coughlin}, S., {Zevin}, M., {et~al.} 2020, \apj, 898, 71,
  \dodoi{10.3847/1538-4357/ab9d85}

\bibitem[{{Broekgaarden} \& {Berger}(2021)}]{broekgaarden:21}
{Broekgaarden}, F.~S., \& {Berger}, E. 2021, \apjl, 920, L13,
  \dodoi{10.3847/2041-8213/ac2832}

\bibitem[{{Brott} {et~al.}(2011){Brott}, {de Mink}, {Cantiello}, {Langer}, {de
  Koter}, {Evans}, {Hunter}, {Trundle}, \& {Vink}}]{brott:11}
{Brott}, I., {de Mink}, S.~E., {Cantiello}, M., {et~al.} 2011, \aap, 530, A115,
  \dodoi{10.1051/0004-6361/201016113}

\bibitem[{{Cantiello} {et~al.}(2007){Cantiello}, {Yoon}, {Langer}, \&
  {Livio}}]{cantiello:2007}
{Cantiello}, M., {Yoon}, S., {Langer}, N., \& {Livio}, M. 2007, \aap, 465, L29

\bibitem[{{Chawla} {et~al.}(2021){Chawla}, {Chatterjee}, {Breivik}, {Krishna
  Moorthy}, {Andrews}, \& {Sanderson}}]{chawla:21}
{Chawla}, C., {Chatterjee}, S., {Breivik}, K., {et~al.} 2021, arXiv e-prints,
  arXiv:2110.05979.
\newblock \doarXiv{2110.05979}

\bibitem[{{Claret} \& {Torres}(2017)}]{claret:17}
{Claret}, A., \& {Torres}, G. 2017, \apj, 849, 18,
  \dodoi{10.3847/1538-4357/aa8770}

\bibitem[{{de Kool}(1990)}]{dekool:1990}
{de Kool}, M. 1990, \apj, 358, 189, \dodoi{10.1086/168974}

\bibitem[{{De Marco} {et~al.}(2011){De Marco}, {Passy}, {Moe}, {Herwig}, {Mac
  Low}, \& {Paxton}}]{demarco:11}
{De Marco}, O., {Passy}, J.-C., {Moe}, M., {et~al.} 2011, \mnras, 411, 2277,
  \dodoi{10.1111/j.1365-2966.2010.17891.x}

\bibitem[{{de Mink} \& {Mandel}(2016)}]{demink:2016}
{de Mink}, S.~E., \& {Mandel}, I. 2016, \mnras, 460, 3545,
  \dodoi{10.1093/mnras/stw1219}

\bibitem[{{Delgado} \& {Thomas}(1981)}]{delgado:81}
{Delgado}, A.~J., \& {Thomas}, H.~C. 1981, \aap, 96, 142

\bibitem[{{Dewi} \& {Tauris}(2000)}]{dewi:2000}
{Dewi}, J.~D.~M., \& {Tauris}, T.~M. 2000, \aap, 360, 1043

\bibitem[{{Dominik} {et~al.}(2012){Dominik}, {Belczynski}, {Fryer}, {Holz},
  {Berti}, {Bulik}, {Mandel}, \& {O'Shaughnessy}}]{dominik:2012}
{Dominik}, M., {Belczynski}, K., {Fryer}, C., {et~al.} 2012, \apj, 759, 52,
  \dodoi{10.1088/0004-637X/759/1/52}

\bibitem[{{Dorn-Wallenstein} {et~al.}(2020){Dorn-Wallenstein}, {Levesque},
  {Neugent}, {Davenport}, {Morris}, \& {Gootkin}}]{dorn-wallenstein:20}
{Dorn-Wallenstein}, T.~Z., {Levesque}, E.~M., {Neugent}, K.~F., {et~al.} 2020,
  \apj, 902, 24, \dodoi{10.3847/1538-4357/abb318}

\bibitem[{{Eggleton}(1983)}]{eggleton:83}
{Eggleton}, P.~P. 1983, \apj, 268, 368, \dodoi{10.1086/160960}

\bibitem[{{El-Badry} {et~al.}(2022){El-Badry}, {Seeburger}, {Jayasinghe},
  {Rix}, {Almada}, {Conroy}, {Price-Whelan}, \& {Burdge}}]{el-badry:2022a}
{El-Badry}, K., {Seeburger}, R., {Jayasinghe}, T., {et~al.} 2022, \mnras, 512,
  5620, \dodoi{10.1093/mnras/stac815}

\bibitem[{Farmer \& Bauer(2018)}]{pymesa}
Farmer, R., \& Bauer, E.~B. 2018, pyMesa, \dodoi{10.5281/zenodo.1205271}

\bibitem[{{Farmer} {et~al.}(2016){Farmer}, {Fields}, {Petermann}, {Dessart},
  {Cantiello}, {Paxton}, \& {Timmes}}]{farmer:16}
{Farmer}, R., {Fields}, C.~E., {Petermann}, I., {et~al.} 2016, \apjs, 227, 22,
  \dodoi{10.3847/1538-4365/227/2/22}

\bibitem[{Farmer {et~al.}(2019)Farmer, Renzo, {de Mink}, Marchant, \&
  Justham}]{farmer:2019}
Farmer, R., Renzo, M., {de Mink}, S.~E., Marchant, P., \& Justham, S. 2019, The
  Astrophysical Journal, 887, 53, \dodoi{10.3847/1538-4357/ab518b}

\bibitem[{{Farrell} {et~al.}(2022){Farrell}, {Groh}, {Meynet}, \&
  {Eldridge}}]{farrell:22}
{Farrell}, E., {Groh}, J.~H., {Meynet}, G., \& {Eldridge}, J.~J. 2022, \mnras,
  512, 4116, \dodoi{10.1093/mnras/stac538}

\bibitem[{{Fragos} {et~al.}(2022){Fragos}, {Andrews}, {Bavera}, {Berry},
  {Coughlin}, {Dotter}, {Giri}, {Kalogera}, {Katsaggelos}, {Kovlakas},
  {Lalvani}, {Misra}, {Srivastava}, {Qin}, {Rocha}, {Roman-Garza}, {Serra},
  {Stahle}, {Sun}, {Teng}, {Trajcevski}, {Hai Tran}, {Xing}, {Zapartas}, \&
  {Zevin}}]{fragos:2022}
{Fragos}, T., {Andrews}, J.~J., {Bavera}, S.~S., {et~al.} 2022, arXiv e-prints,
  arXiv:2202.05892.
\newblock \doarXiv{2202.05892}

\bibitem[{Fryer {et~al.}(2012)Fryer, Belczynski, Wiktorowicz, Dominik,
  Kalogera, \& Holz}]{fryer:2012}
Fryer, C.~L., Belczynski, K., Wiktorowicz, G., {et~al.} 2012, The Astrophysical
  Journal, 749, 91, \dodoi{10.1088/0004-637X/749/1/91}

\bibitem[{{Fryer} {et~al.}(2022){Fryer}, {Olejak}, \&
  {Belczynski}}]{fryer:2022}
{Fryer}, C.~L., {Olejak}, A., \& {Belczynski}, K. 2022, \apj, 931, 94,
  \dodoi{10.3847/1538-4357/ac6ac9}

\bibitem[{Gagnier {et~al.}(2019)Gagnier, Rieutord, Charbonnel, Putigny, \&
  Espinosa~Lara}]{gagnier:2019}
Gagnier, D., Rieutord, M., Charbonnel, C., Putigny, B., \& Espinosa~Lara, F.
  2019, Astronomy and Astrophysics, 625, 1, \dodoi{10.1051/0004-6361/201832581}

\bibitem[{{G{\"o}tberg} {et~al.}(2018){G{\"o}tberg}, {de Mink}, {Groh},
  {Kupfer}, {Crowther}, {Zapartas}, \& {Renzo}}]{gotberg:2018}
{G{\"o}tberg}, Y., {de Mink}, S.~E., {Groh}, J.~H., {et~al.} 2018, \aap, 615,
  A78, \dodoi{10.1051/0004-6361/201732274}

\bibitem[{{Han} {et~al.}(1995){Han}, {Podsiadlowski}, \& {Eggleton}}]{han:95}
{Han}, Z., {Podsiadlowski}, P., \& {Eggleton}, P.~P. 1995, \mnras, 272, 800,
  \dodoi{10.1093/mnras/272.4.800}

\bibitem[{{Heger} {et~al.}(2000){Heger}, {Langer}, \& {Woosley}}]{heger:2000}
{Heger}, A., {Langer}, N., \& {Woosley}, S.~E. 2000, \apj, 528, 368

\bibitem[{{Hellings}(1983)}]{hellings:1983}
{Hellings}, P. 1983, \apss, 96, 37, \dodoi{10.1007/BF00661941}

\bibitem[{{Hellings}(1984)}]{hellings:1984}
---. 1984, \apss, 104, 83, \dodoi{10.1007/BF00653994}

\bibitem[{Hirai {et~al.}(2018)Hirai, Podsiadlowski, \& Yamada}]{hirai:2018}
Hirai, R., Podsiadlowski, P., \& Yamada, S. 2018, The Astrophysical Journal,
  864, 119, \dodoi{10.3847/1538-4357/aad6a0}

\bibitem[{{Hjellming} \& {Webbink}(1987)}]{hjellming:1987}
{Hjellming}, M.~S., \& {Webbink}, R.~F. 1987, \apj, 318, 794,
  \dodoi{10.1086/165412}

\bibitem[{{Hunter}(2007)}]{matplotlib}
{Hunter}, J.~D. 2007, Computing in Science and Engineering, 9, 90,
  \dodoi{10.1109/MCSE.2007.55}

\bibitem[{{Hurley} {et~al.}(2002){Hurley}, {Tout}, \& {Pols}}]{hurley:2002}
{Hurley}, J.~R., {Tout}, C.~A., \& {Pols}, O.~R. 2002, \mnras, 329, 897,
  \dodoi{10.1046/j.1365-8711.2002.05038.x}

\bibitem[{Ivanova {et~al.}(2020)Ivanova, Justham, \& Ricker}]{ivanova:2020}
Ivanova, N., Justham, S., \& Ricker, P. 2020, Common Envelope Evolution,
  2514-3433 (IOP Publishing), \dodoi{10.1088/2514-3433/abb6f0}

\bibitem[{Ivanova {et~al.}(2013)Ivanova, Justham, Chen, De~Marco, Fryer,
  Gaburov, Ge, Glebbeek, Han, Li, Lu, Marsh, Podsiadlowski, Potter, Soker,
  Taam, Tauris, {van den Heuvel}, \& Webbink}]{ivanova:2013}
Ivanova, N., Justham, S., Chen, X., {et~al.} 2013, The Astronomy and
  Astrophysics Review, 21, 59, \dodoi{10.1007/s00159-013-0059-2}

\bibitem[{{Jermyn} {et~al.}(2022){Jermyn}, {Bauer}, {Schwab}, {Farmer}, {Ball},
  {Bellinger}, {Dotter}, {Joyce}, {Marchant}, {Mombarg}, {Wolf}, {Wong},
  {Cinquegrana}, {Farrell}, {Smolec}, {Thoul}, {Cantiello}, {Herwig}, {Toloza},
  {Bildsten}, {Townsend}, \& {Timmes}}]{jermyn:2022}
{Jermyn}, A.~S., {Bauer}, E.~B., {Schwab}, J., {et~al.} 2022, arXiv e-prints,
  arXiv:2208.03651.
\newblock \doarXiv{2208.03651}

\bibitem[{Kalogera(1996)}]{kalogera:1996}
Kalogera, V. 1996, \apj, 471, 352, \dodoi{10.1086/177974}

\bibitem[{{Kippenhahn} \& {Weigert}(1967)}]{kippenhahn:1967}
{Kippenhahn}, R., \& {Weigert}, A. 1967, \zap, 65, 251

\bibitem[{{Klencki} {et~al.}(2022){Klencki}, {Istrate}, {Nelemans}, \&
  {Pols}}]{klencki:2021}
{Klencki}, J., {Istrate}, A., {Nelemans}, G., \& {Pols}, O. 2022, \aap, 662,
  A56, \dodoi{10.1051/0004-6361/202142701}

\bibitem[{{Klencki} {et~al.}(2021){Klencki}, {Nelemans}, {Istrate}, \&
  {Chruslinska}}]{klencki:2020}
{Klencki}, J., {Nelemans}, G., {Istrate}, A.~G., \& {Chruslinska}, M. 2021,
  \aap, 645, A54, \dodoi{10.1051/0004-6361/202038707}

\bibitem[{Korol {et~al.}(2017)Korol, Rossi, Groot, Nelemans, Toonen, \&
  Brown}]{korol:2017}
Korol, V., Rossi, E.~M., Groot, P.~J., {et~al.} 2017, Monthly Notices of the
  Royal Astronomical Society, 470, 1894, \dodoi{10.1093/mnras/stx1285}

\bibitem[{{Kremer} {et~al.}(2017){Kremer}, {Breivik}, {Larson}, \&
  {Kalogera}}]{kremer:2017}
{Kremer}, K., {Breivik}, K., {Larson}, S.~L., \& {Kalogera}, V. 2017, \apj,
  846, 95, \dodoi{10.3847/1538-4357/aa8557}

\bibitem[{{Kruckow} {et~al.}(2018){Kruckow}, {Tauris}, {Langer}, {Kramer}, \&
  {Izzard}}]{kruckow:18}
{Kruckow}, M.~U., {Tauris}, T.~M., {Langer}, N., {Kramer}, M., \& {Izzard},
  R.~G. 2018, \mnras, 481, 1908, \dodoi{10.1093/mnras/sty2190}

\bibitem[{Langer(1998)}]{langer:1998}
Langer, N. 1998, Astronomy and Astrophysics, 329, 551

\bibitem[{Laplace {et~al.}(2020)Laplace, G{\"o}tberg, De~Mink, Justham, \&
  Farmer}]{laplace:2020}
Laplace, E., G{\"o}tberg, Y., De~Mink, S.~E., Justham, S., \& Farmer, R. 2020,
  Astronomy and Astrophysics, 637, A6, \dodoi{10.1051/0004-6361/201937300}

\bibitem[{{Laplace} {et~al.}(2021){Laplace}, {Justham}, {Renzo}, {G{\"o}tberg},
  {Farmer}, {Vartanyan}, \& {de Mink}}]{laplace:2021}
{Laplace}, E., {Justham}, S., {Renzo}, M., {et~al.} 2021, \aap, 656, A58,
  \dodoi{10.1051/0004-6361/202140506}

\bibitem[{{Lau} {et~al.}(2022{\natexlab{a}}){Lau}, {Hirai},
  {Gonz{\'a}lez-Bol{\'\i}var}, {Price}, {De Marco}, \& {Mandel}}]{lau:2022}
{Lau}, M. Y.~M., {Hirai}, R., {Gonz{\'a}lez-Bol{\'\i}var}, M., {et~al.}
  2022{\natexlab{a}}, \mnras, \dodoi{10.1093/mnras/stac049}

\bibitem[{{Lau} {et~al.}(2022{\natexlab{b}}){Lau}, {Hirai}, {Price}, \&
  {Mandel}}]{lau:22}
{Lau}, M. Y.~M., {Hirai}, R., {Price}, D.~J., \& {Mandel}, I.
  2022{\natexlab{b}}, arXiv e-prints, arXiv:2206.06411.
\newblock \doarXiv{2206.06411}

\bibitem[{{Law-Smith} {et~al.}(2020){Law-Smith}, {Everson}, {Ramirez-Ruiz}, {de
  Mink}, {van Son}, {G{\"o}tberg}, {Zellmann}, {Vigna-G{\'o}mez}, {Renzo},
  {Wu}, {Schr{\o}der}, {Foley}, \& {Hutchinson-Smith}}]{law-smith:2020}
{Law-Smith}, J. A.~P., {Everson}, R.~W., {Ramirez-Ruiz}, E., {et~al.} 2020,
  arXiv e-prints, arXiv:2011.06630.
\newblock \doarXiv{2011.06630}

\bibitem[{{Lubow} \& {Shu}(1975)}]{lubow:1975}
{Lubow}, S.~H., \& {Shu}, F.~H. 1975, \apj, 198, 383, \dodoi{10.1086/153614}

\bibitem[{{Luger} {et~al.}(2021){Luger}, {Bedell}, {Foreman-Mackey},
  {Crossfield}, {Zhao}, \& {Hogg}}]{showyourwork}
{Luger}, R., {Bedell}, M., {Foreman-Mackey}, D., {et~al.} 2021, arXiv e-prints,
  arXiv:2110.06271.
\newblock \doarXiv{2110.06271}

\bibitem[{{Maeder} \& {Meynet}(2000)}]{maeder:00}
{Maeder}, A., \& {Meynet}, G. 2000, \araa, 38, 143,
  \dodoi{10.1146/annurev.astro.38.1.143}

\bibitem[{{Marchant} {et~al.}(2021){Marchant}, {Pappas}, {Gallegos-Garcia},
  {Berry}, {Taam}, {Kalogera}, \& {Podsiadlowski}}]{marchant:2021}
{Marchant}, P., {Pappas}, K. M.~W., {Gallegos-Garcia}, M., {et~al.} 2021, \aap,
  650, A107, \dodoi{10.1051/0004-6361/202039992}

\bibitem[{{Moravveji} {et~al.}(2016){Moravveji}, {Townsend}, {Aerts}, \&
  {Mathis}}]{moravveji:16}
{Moravveji}, E., {Townsend}, R.~H.~D., {Aerts}, C., \& {Mathis}, S. 2016, \apj,
  823, 130, \dodoi{10.3847/0004-637X/823/2/130}

\bibitem[{{Morton}(1960)}]{morton:60}
{Morton}, D.~C. 1960, \apj, 132, 146, \dodoi{10.1086/146908}

\bibitem[{M{\"u}ller \& Vink(2014)}]{muller:2014}
M{\"u}ller, P.~E., \& Vink, J.~S. 2014, Astronomy and Astrophysics, 564, 1,
  \dodoi{10.1051/0004-6361/201323031}

\bibitem[{{Nandez} {et~al.}(2014){Nandez}, {Ivanova}, \&
  {Lombardi}}]{nandez:14}
{Nandez}, J.~L.~A., {Ivanova}, N., \& {Lombardi}, J.~C., J. 2014, \apj, 786,
  39, \dodoi{10.1088/0004-637X/786/1/39}

\bibitem[{Neo {et~al.}(1977)Neo, Miyaji, Nomoto, \& Sugimoto}]{neo:1977}
Neo, S., Miyaji, S., Nomoto, K., \& Sugimoto, D. 1977, \textbackslash pasj, 29,
  249

\bibitem[{{O'Connor} \& {Ott}(2011)}]{oconnor:11}
{O'Connor}, E., \& {Ott}, C.~D. 2011, \apj, 730, 70

\bibitem[{{Ogata} {et~al.}(2021){Ogata}, {Hirai}, \& {Hijikawa}}]{ogata:2021}
{Ogata}, M., {Hirai}, R., \& {Hijikawa}, K. 2021, \mnras, 505, 2485,
  \dodoi{10.1093/mnras/stab1439}

\bibitem[{Packet(1981)}]{packet:1981}
Packet, W. 1981, A\&A, 1, \dodoi{10.1017/CBO9781107415324.004}

\bibitem[{{Paczy{\'n}ski}(1971)}]{paczynski:1971}
{Paczy{\'n}ski}, B. 1971, \araa, 9, 183,
  \dodoi{10.1146/annurev.aa.09.090171.001151}

\bibitem[{{Paczynski}(1976)}]{paczynski:1976}
{Paczynski}, B. 1976, in Structure and Evolution of Close Binary Systems, ed.
  P.~{Eggleton}, S.~{Mitton}, \& J.~{Whelan}, Vol.~73, 75

\bibitem[{{Patton} \& {Sukhbold}(2020)}]{patton:2020}
{Patton}, R.~A., \& {Sukhbold}, T. 2020, \mnras, 499, 2803,
  \dodoi{10.1093/mnras/staa3029}

\bibitem[{{Patton} {et~al.}(2022){Patton}, {Sukhbold}, \&
  {Eldridge}}]{patton:22}
{Patton}, R.~A., {Sukhbold}, T., \& {Eldridge}, J.~J. 2022, \mnras, 511, 903,
  \dodoi{10.1093/mnras/stab3797}

\bibitem[{Pavlovskii {et~al.}(2017)Pavlovskii, Ivanova, Belczynski, \&
  Van}]{pavlovskii:2017}
Pavlovskii, K., Ivanova, N., Belczynski, K., \& Van, K.~X. 2017, Monthly
  Notices of the Royal Astronomical Society, 465, 2092,
  \dodoi{10.1093/mnras/stw2786}

\bibitem[{{Paxton} {et~al.}(2011){Paxton}, {Bildsten}, {Dotter}, {Herwig},
  {Lesaffre}, \& {Timmes}}]{paxton:2011}
{Paxton}, B., {Bildsten}, L., {Dotter}, A., {et~al.} 2011, \apjs, 192, 3,
  \dodoi{10.1088/0067-0049/192/1/3}

\bibitem[{{Paxton} {et~al.}(2013){Paxton}, {Cantiello}, {Arras}, {Bildsten},
  {Brown}, {Dotter}, {Mankovich}, {Montgomery}, {Stello}, {Timmes}, \&
  {Townsend}}]{paxton:2013}
{Paxton}, B., {Cantiello}, M., {Arras}, P., {et~al.} 2013, \apjs, 208, 4,
  \dodoi{10.1088/0067-0049/208/1/4}

\bibitem[{{Paxton} {et~al.}(2015){Paxton}, {Marchant}, {Schwab}, {Bauer},
  {Bildsten}, {Cantiello}, {Dessart}, {Farmer}, {Hu}, {Langer}, {Townsend},
  {Townsley}, \& {Timmes}}]{paxton:2015}
{Paxton}, B., {Marchant}, P., {Schwab}, J., {et~al.} 2015, \apjs, 220, 15,
  \dodoi{10.1088/0067-0049/220/1/15}

\bibitem[{{Paxton} {et~al.}(2018){Paxton}, {Schwab}, {Bauer}, {Bildsten},
  {Blinnikov}, {Duffell}, {Farmer}, {Goldberg}, {Marchant}, {Sorokina},
  {Thoul}, {Townsend}, \& {Timmes}}]{paxton:2018}
{Paxton}, B., {Schwab}, J., {Bauer}, E.~B., {et~al.} 2018, \apjs, 234, 34,
  \dodoi{10.3847/1538-4365/aaa5a8}

\bibitem[{{Paxton} {et~al.}(2019){Paxton}, {Smolec}, {Schwab}, {Gautschy},
  {Bildsten}, {Cantiello}, {Dotter}, {Farmer}, {Goldberg}, {Jermyn}, {Kanbur},
  {Marchant}, {Thoul}, {Townsend}, {Wolf}, {Zhang}, \& {Timmes}}]{paxton:2019}
{Paxton}, B., {Smolec}, R., {Schwab}, J., {et~al.} 2019, \apjs, 243, 10,
  \dodoi{10.3847/1538-4365/ab2241}

\bibitem[{{Pejcha} {et~al.}(2017){Pejcha}, {Metzger}, {Tyles}, \&
  {Tomida}}]{pejcha:17}
{Pejcha}, O., {Metzger}, B.~D., {Tyles}, J.~G., \& {Tomida}, K. 2017, \apj,
  850, 59, \dodoi{10.3847/1538-4357/aa95b9}

\bibitem[{Perez \& Granger(2007)}]{ipython}
Perez, F., \& Granger, B.~E. 2007, Computing in Science Engineering, 9, 21,
  \dodoi{10.1109/MCSE.2007.53}

\bibitem[{{Ram{\'{\i}}rez-Agudelo} {et~al.}(2015){Ram{\'{\i}}rez-Agudelo},
  {Sana}, {de Mink}, {H{\'e}nault-Brunet}, {de Koter}, {Langer}, {Tramper},
  {Gr{\"a}fener}, {Evans}, {Vink}, {Dufton}, \&
  {Taylor}}]{ramirez-agudelo:2015}
{Ram{\'{\i}}rez-Agudelo}, O.~H., {Sana}, H., {de Mink}, S.~E., {et~al.} 2015,
  \aap, 580, A92, \dodoi{10.1051/0004-6361/201425424}

\bibitem[{Renzo {et~al.}(2020)Renzo, Farmer, Justham, Mink, \&
  Marchant}]{renzo:2020ppi_conv}
Renzo, M., Farmer, R.~J., Justham, S., Mink, S. E.~D., \& Marchant, P. 2020,
  Monthly Notices of the Royal Astronomical Society, 4341, 4333,
  \dodoi{10.1093/mnras/staa549}

\bibitem[{{Renzo} \& {G{\"o}tberg}(2021)}]{renzo:2021zoph}
{Renzo}, M., \& {G{\"o}tberg}, Y. 2021, \apj, 923, 277,
  \dodoi{10.3847/1538-4357/ac29c5}

\bibitem[{{Renzo} {et~al.}(2022){Renzo}, {Hendriks}, {van Son}, \&
  {Farmer}}]{renzo:2022}
{Renzo}, M., {Hendriks}, D.~D., {van Son}, L.~A.~C., \& {Farmer}, R. 2022,
  Research Notes of the American Astronomical Society, 6, 25,
  \dodoi{10.3847/2515-5172/ac503e}

\bibitem[{Renzo {et~al.}(2017)Renzo, Ott, Shore, \& De~Mink}]{renzo:2017}
Renzo, M., Ott, C., Shore, S., \& De~Mink, S. 2017, Astronomy and Astrophysics,
  603, \dodoi{10.1051/0004-6361/201730698}

\bibitem[{Renzo {et~al.}(2019)Renzo, Zapartas, {de Mink}, G{\"o}tberg, Justham,
  Farmer, Izzard, Toonen, \& Sana}]{renzo:2019walk}
Renzo, M., Zapartas, E., {de Mink}, S.~E., {et~al.} 2019, Astronomy \&
  Astrophysics, 66, 1, \dodoi{10.1051/0004-6361/201833297}

\bibitem[{{Renzo} {et~al.}(2021){Renzo}, {Callister}, {Chatziioannou}, {van
  Son}, {Mingarelli}, {Cantiello}, {Ford}, {McKernan}, \&
  {Ashton}}]{renzo:21gwce}
{Renzo}, M., {Callister}, T., {Chatziioannou}, K., {et~al.} 2021, \apj, 919,
  128, \dodoi{10.3847/1538-4357/ac1110}

\bibitem[{{Sen} {et~al.}(2022){Sen}, {Langer}, {Marchant}, {Menon}, {de Mink},
  {Schootemeijer}, {Sch{\"u}rmann}, {Mahy}, {Hastings}, {Nathaniel}, {Sana},
  {Wang}, \& {Xu}}]{sen:2022}
{Sen}, K., {Langer}, N., {Marchant}, P., {et~al.} 2022, \aap, 659, A98,
  \dodoi{10.1051/0004-6361/202142574}

\bibitem[{Sravan {et~al.}(2019)Sravan, Marchant, \& Kalogera}]{sravan:2019}
Sravan, N., Marchant, P., \& Kalogera, V. 2019, The Astrophysical Journal, 885,
  130, \dodoi{10.3847/1538-4357/ab4ad7}

\bibitem[{{Tauris} \& {Dewi}(2001)}]{tauris:01}
{Tauris}, T.~M., \& {Dewi}, J.~D.~M. 2001, \aap, 369, 170,
  \dodoi{10.1051/0004-6361:20010099}

\bibitem[{{Tauris} {et~al.}(2017){Tauris}, {Kramer}, {Freire}, {Wex}, {Janka},
  {Langer}, {Podsiadlowski}, {Bozzo}, {Chaty}, {Kruckow}, {van den Heuvel},
  {Antoniadis}, {Breton}, \& {Champion}}]{tauris:2017}
{Tauris}, T.~M., {Kramer}, M., {Freire}, P.~C.~C., {et~al.} 2017, \apj, 846,
  170, \dodoi{10.3847/1538-4357/aa7e89}

\bibitem[{Tauris \& Takens(1998)}]{tauris:1998}
Tauris, T.~M.~M., \& Takens, R.~J.~J. 1998, Astronomy and Astrophysics, 1059,
  1047

\bibitem[{{Thiele} {et~al.}(2021){Thiele}, {Breivik}, \&
  {Sanderson}}]{thiele:21}
{Thiele}, S., {Breivik}, K., \& {Sanderson}, R.~E. 2021, arXiv e-prints,
  arXiv:2111.13700.
\newblock \doarXiv{2111.13700}

\bibitem[{{Tutukov} \& {Yungelson}(1993)}]{tutukov:93}
{Tutukov}, A.~V., \& {Yungelson}, L.~R. 1993, \mnras, 260, 675,
  \dodoi{10.1093/mnras/260.3.675}

\bibitem[{{van den Heuvel} {et~al.}(2017){van den Heuvel}, {Portegies Zwart},
  \& {de Mink}}]{vandenheuvel:2017}
{van den Heuvel}, E.~P.~J., {Portegies Zwart}, S.~F., \& {de Mink}, S.~E. 2017,
  \mnras, 471, 4256, \dodoi{10.1093/mnras/stx1430}

\bibitem[{{van der Walt} {et~al.}(2011){van der Walt}, {Colbert}, \&
  {Varoquaux}}]{numpy}
{van der Walt}, S., {Colbert}, S.~C., \& {Varoquaux}, G. 2011, Computing in
  Science and Engineering, 13, 22, \dodoi{10.1109/MCSE.2011.37}

\bibitem[{{van Son} {et~al.}(2022){van Son}, {de Mink}, {Callister}, {Justham},
  {Renzo}, {Wagg}, {Broekgaarden}, {Kummer}, {Pakmor}, \&
  {Mandel}}]{vanson:2021}
{van Son}, L.~A.~C., {de Mink}, S.~E., {Callister}, T., {et~al.} 2022, \apj,
  931, 17, \dodoi{10.3847/1538-4357/ac64a3}

\bibitem[{{Vigna-G{\'o}mez} {et~al.}(2018){Vigna-G{\'o}mez}, Neijssel,
  Stevenson, Barrett, Belczynski, Justham, {de~Mink}, M{\"u}ller,
  Podsiadlowski, Renzo, Sz{\'e}csi, \& Mandel}]{vigna-gomez:2018}
{Vigna-G{\'o}mez}, A., Neijssel, C.~J., Stevenson, S., {et~al.} 2018, Monthly
  Notices of the Royal Astronomical Society, 481, 4009,
  \dodoi{10.1093/mnras/sty2463}

\bibitem[{{Vigna-G{\'o}mez} {et~al.}(2020){Vigna-G{\'o}mez}, {MacLeod},
  {Neijssel}, {Broekgaarden}, {Justham}, {Howitt}, {de Mink}, {Vinciguerra}, \&
  {Mandel}}]{vigna-gomez:2020}
{Vigna-G{\'o}mez}, A., {MacLeod}, M., {Neijssel}, C.~J., {et~al.} 2020, \pasa,
  37, e038, \dodoi{10.1017/pasa.2020.31}

\bibitem[{Virtanen {et~al.}(2020)Virtanen, Gommers, Oliphant, Haberland, Reddy,
  Cournapeau, Burovski, Peterson, Weckesser, Bright, {van der Walt}, Brett,
  Wilson, Millman, Mayorov, Nelson, Jones, Kern, Larson, Carey, Polat, Feng,
  Moore, {VanderPlas}, Laxalde, Perktold, Cimrman, Henriksen, Quintero, Harris,
  Archibald, Ribeiro, Pedregosa, {van Mulbregt}, \& {SciPy 1.0
  Contributors}}]{scipy}
Virtanen, P., Gommers, R., Oliphant, T.~E., {et~al.} 2020, Nature Methods, 17,
  261, \dodoi{10.1038/s41592-019-0686-2}

\bibitem[{Walmswell {et~al.}(2015)Walmswell, Tout, \&
  Eldridge}]{walmswell:2015}
Walmswell, J., Tout, C.~A., \& Eldridge, J.~J. 2015, \mnras, 447, 2951,
  \dodoi{10.1093/mnras/stu2666}

\bibitem[{{Wang} {et~al.}(2020){Wang}, {Langer}, {Schootemeijer}, {Castro},
  {Adscheid}, {Marchant}, \& {Hastings}}]{wang:2020}
{Wang}, C., {Langer}, N., {Schootemeijer}, A., {et~al.} 2020, \apjl, 888, L12,
  \dodoi{10.3847/2041-8213/ab6171}

\bibitem[{Wang {et~al.}(2021)Wang, Gies, Peters, G{\"o}tberg, Chojnowski,
  Lester, \& Howell}]{wang:2021a}
Wang, L., Gies, D.~R., Peters, G.~J., {et~al.} 2021.
\newblock \doarXiv{2103.13642}

\bibitem[{{Webbink}(1984)}]{webbink:1984}
{Webbink}, R.~F. 1984, \apj, 277, 355, \dodoi{10.1086/161701}

\bibitem[{{Yoon} \& {Langer}(2005)}]{yoon:2005}
{Yoon}, S.~C., \& {Langer}, N. 2005, \aap, 443, 643,
  \dodoi{10.1051/0004-6361:20054030}

\bibitem[{{Zapartas} {et~al.}(2017){Zapartas}, {de Mink}, {Izzard}, {Yoon},
  {Badenes}, {G{\"o}tberg}, {de Koter}, {Neijssel}, {Renzo}, {Schootemeijer},
  \& {Shrotriya}}]{zapartas:2017}
{Zapartas}, E., {de Mink}, S.~E., {Izzard}, R.~G., {et~al.} 2017, \aap, 601,
  A29, \dodoi{10.1051/0004-6361/201629685}

\bibitem[{{Zapartas} {et~al.}(2021){Zapartas}, {Renzo}, {Fragos}, {Dotter},
  {Andrews}, {Bavera}, {Coughlin}, {Misra}, {Kovlakas}, {Rom{\'a}n-Garza},
  {Serra}, {Qin}, {Rocha}, {Tran}, \& {Xing}}]{zapartas:21b}
{Zapartas}, E., {Renzo}, M., {Fragos}, T., {et~al.} 2021, \aap, 656, L19,
  \dodoi{10.1051/0004-6361/202141506}

\bibitem[{{Zorotovic} {et~al.}(2010){Zorotovic}, {Schreiber}, {G{\"a}nsicke},
  \& {Nebot G{\'o}mez-Mor{\'a}n}}]{zorotovic:2010}
{Zorotovic}, M., {Schreiber}, M.~R., {G{\"a}nsicke}, B.~T., \& {Nebot
  G{\'o}mez-Mor{\'a}n}, A. 2010, \aap, 520, A86,
  \dodoi{10.1051/0004-6361/200913658}

\end{thebibliography}

\appendix
\section{Impact of core-envelope boundary and rotation on the binding  energy profile}
\label{sec:toy_models}

In this appendix, we introduce our ``engineered'' models and
illustrate with examples how the
envelope binding energy depends on the CEB
region
(\Secref{sec:eng_examples}) and on the initial rotation rate of the
star (\Secref{sec:rot_examples}). Both can be significantly modified
by accretion during the first RLOF.

\begin{figure}[hbtp]
  \includegraphics[width=0.5\textwidth]{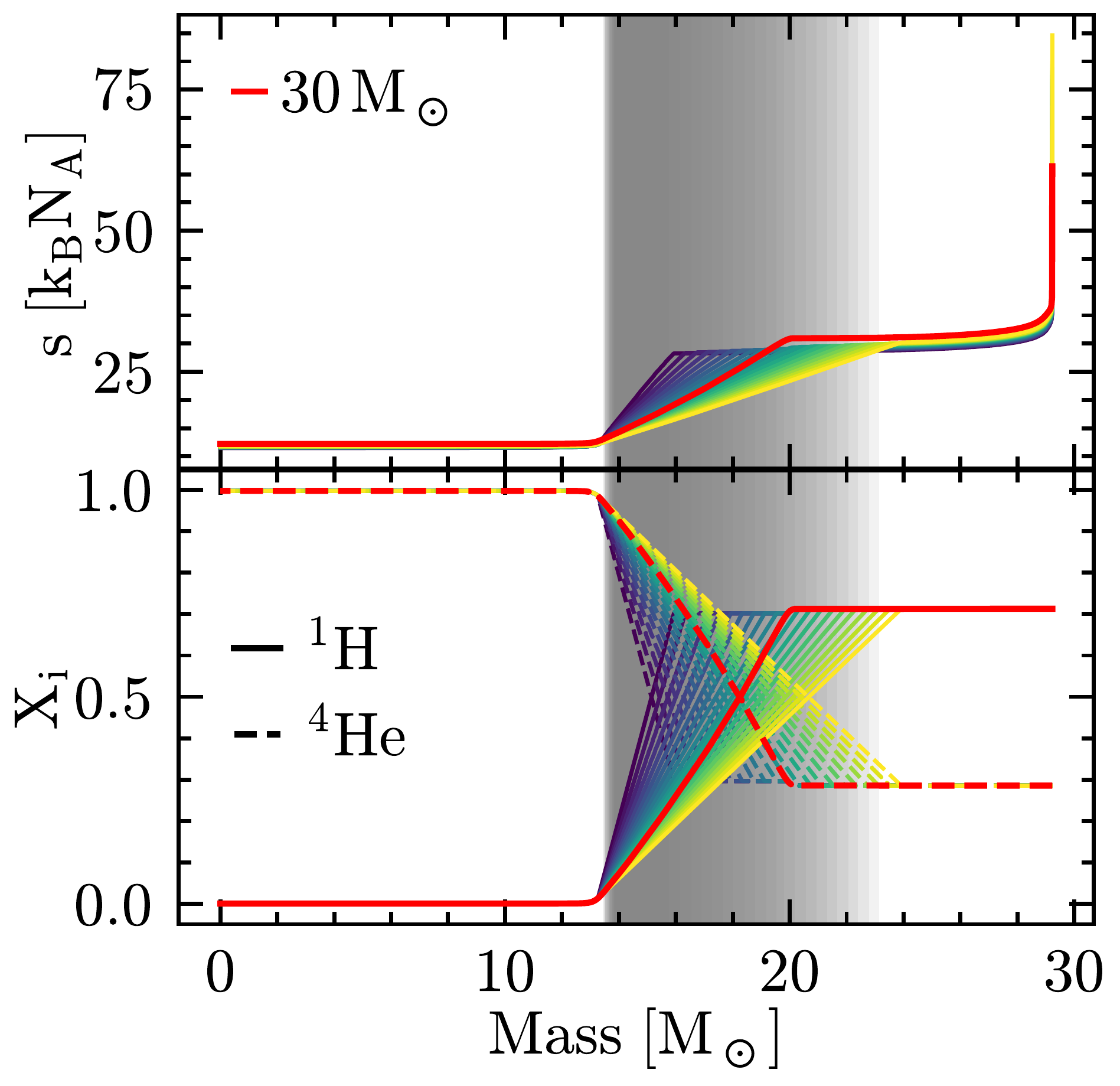}
  \caption{TAMS entropy profile of a single 30\,$M_\odot$ star (red)
    and engineered models where we artificially modify the CEB region
    (gray shaded area partially overlapping for multiple models,
    increasing CEB size from blue to yellow). The CEB for a single
  $30\,M_\odot$ star has a mass thickness of $5.81\,M_\odot$ at TAMS,
  while the engineered models span the range $\sim3-9.4\,M_\odot$.}
  \label{fig:engineered_TAMS}
  \script{engineered_TAMS.py}
\end{figure}

\Figref{fig:engineered_TAMS} shows an example grid of ``engineered
stars'' of $20\,M_\odot$, similar to \Figref{fig:TAMS_profiles}.
Starting from a non-rotating single star at TAMS (e.g., red model in
\Figref{fig:engineered_TAMS}), we modify the CEB specific entropy (s)
which controls the thermal properties of the gas, and its H, and He
profiles -- but do not change the mass fractions of other elements.
Specifically, we keep the same inner and outer profiles, but impose a
linear connection from the outer boundary of the H-depleted core to a
mass coordinate which we specify as a parameter (see
\Figref{fig:engineered_TAMS} and \Figref{fig:TAMS_profiles}). We let \mesa\ relax
the TAMS profiles to the desired entropy and composition profiles and
then recover gravothermal and hydrostatic equilibrium, and then evolve
until either carbon depletion or when the photospheric radius of these
models first exceeds
$1000\,R_\odot$.

\subsection{Steepness of the core-envelope boundary}
\label{sec:eng_examples}

\begin{figure}[htbp]
  \centering
  \includegraphics[width=0.5\textwidth]{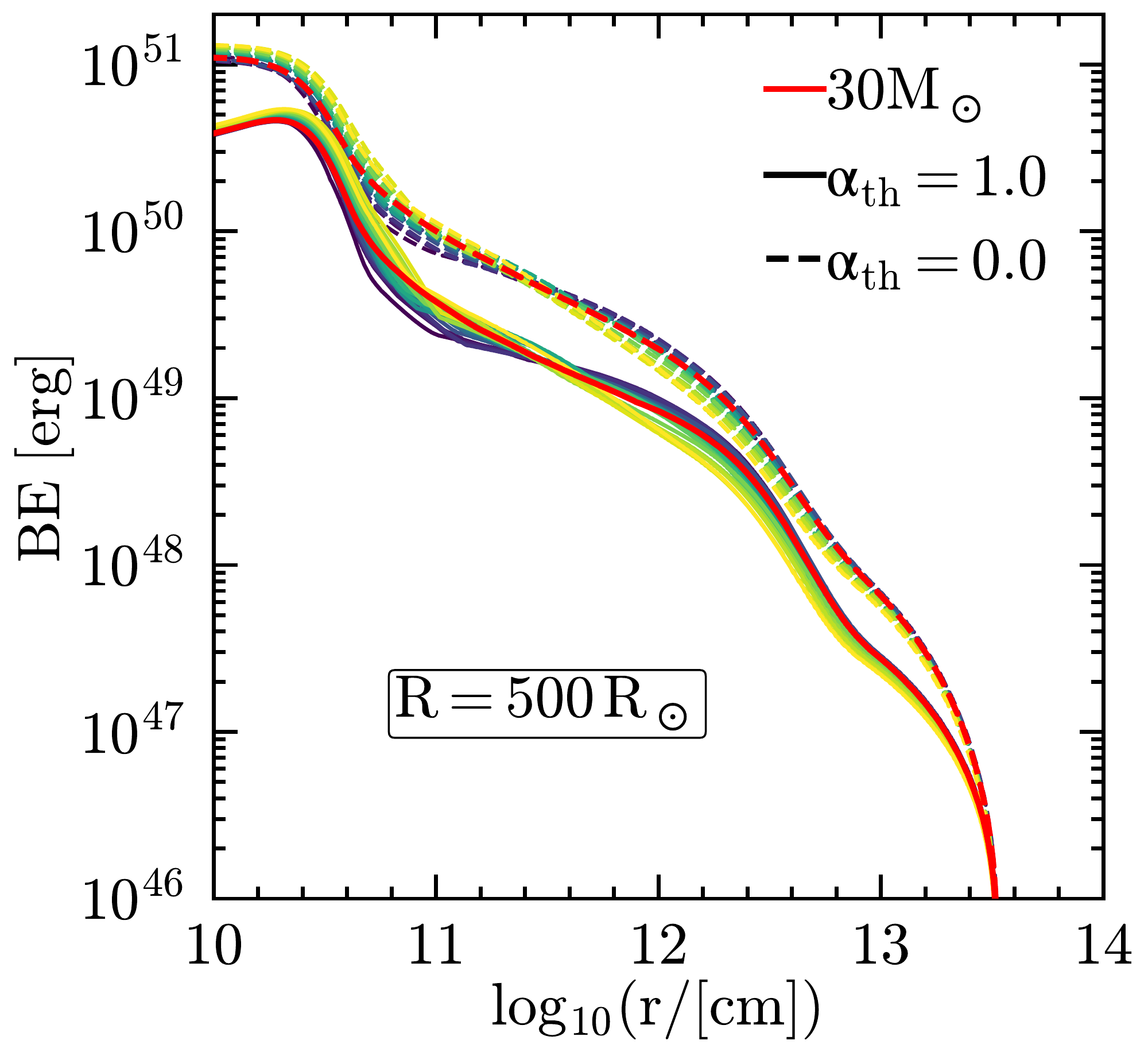}
  \caption{The structure of the CEB at the end of the main sequence
    impacts the envelope binding energy profile throughout the
    remaining evolution. Dashed lines show the gravitational
    contribution only, while solid lines include the contribution of
    the internal energy. The red lines show a $30\,M_\odot$,
    non-rotating, $Z=0.0019$ model compared to ``engineered'' models
    of the same mass (see \Figref{fig:engineered_TAMS},
    increasing CEB size from blue to yellow), but
    artificially imposed profile at TAMS (other colors, see text). The
    top (bottom) axis indicates mass coordinate (radius). We compare
    the models when they first reach $500\,R_\odot$.}
  \label{fig:toy_models_example}
  \script{toy_models_example.py}
\end{figure}

\Figref{fig:toy_models_example} shows a comparison of the
gravitational and binding energy profiles of a $30\,M_\odot$ single
star (red solid line) to
``engineered'' models, when stars first
reach radius $R=500\,R_\odot$.
\Figref{fig:toy_models_example} shows that
binding energy depends on the structure of the CEB region. In single
stars, the CEB is determined by the extent of the convective boundary
mixing and the recession in mass coordinate of the
convective core. In \Figref{fig:toy_models_example}, lines of
different colors show a trend with shallower entropy and composition
profiles at TAMS (lighter curves in \Figref{fig:engineered_TAMS})
evolving into more bound inner envelopes (larger
binding energy inside $\log_{10}(r/\mathrm{cm})\lesssim 11.5$), and
vice versa.

\subsection{Rotation}
\label{sec:rot_examples}

Mass transfer through RLOF also spins up the accreting star, often to
critical rotation\footnote{At critical rotation, the centrifugal force
  balances the gravitational pull at the equator, corresponding to
  critical angular frequency
  $\omega_\mathrm{crit}=\sqrt{(1-L/L_\mathrm{Edd})GM/R^3}$, with
  $L_\mathrm{Edd}$ the Eddington luminosity, and $L$ the stellar
  luminosity.} \citep[e.g.,][]{lubow:1975, packet:1981,
  cantiello:2007, renzo:2021zoph}. To illustrate the
  impact of rotation, it is worth considering the CEB
 region and envelope structure
 of single-star models rotating since
birth, although spinning up a star late during its
main-sequence evolution has different structural consequences than
natal rotation (see \citealt{renzo:2021zoph}).

\begin{figure}[htbp]
  \centering
  \includegraphics[width=0.5\textwidth]{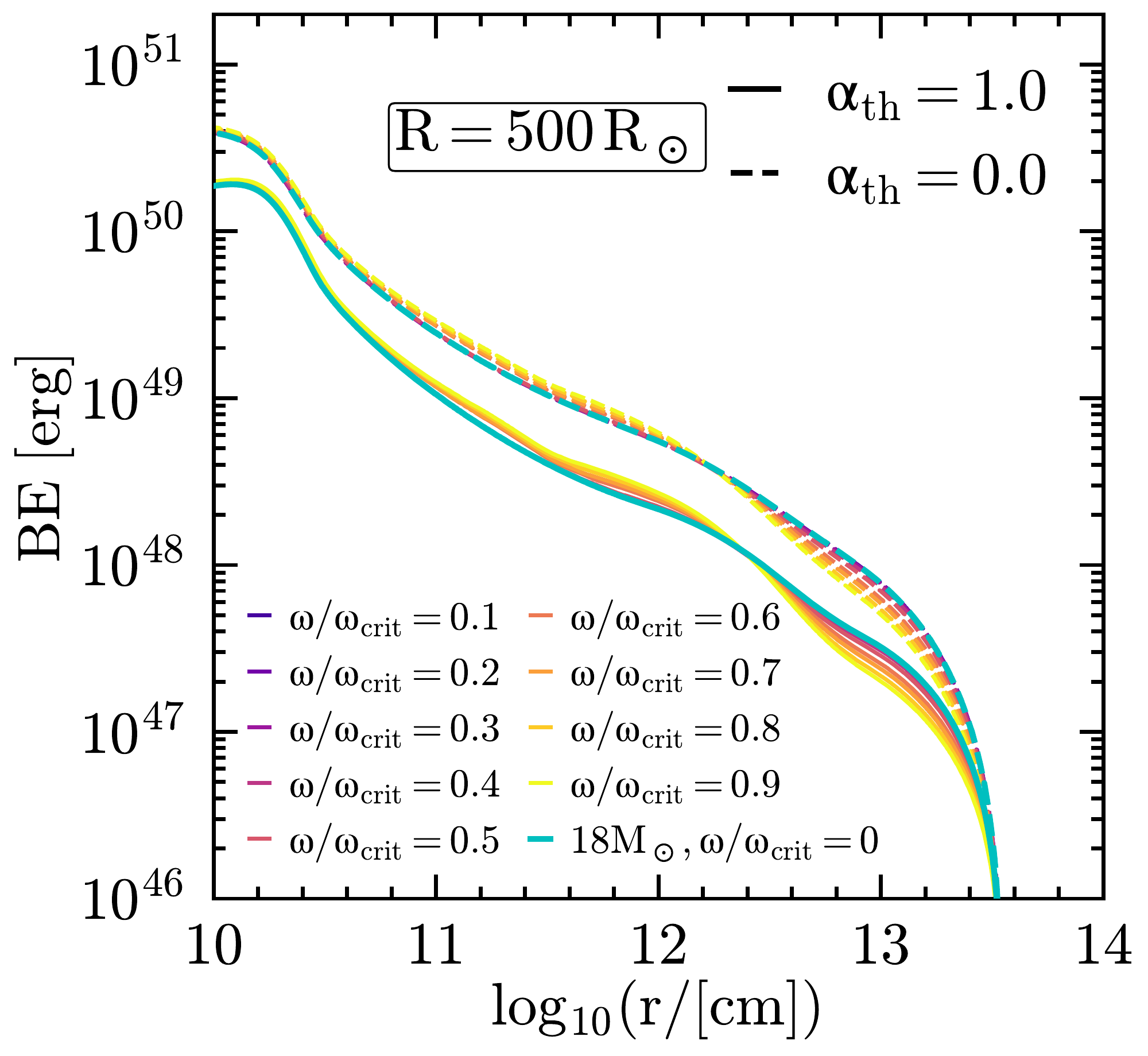}
  \caption{Same as \Figref{fig:toy_models_example} but comparing
    single stars differing by their initial rotation rate.}
  \label{fig:rotation_models_example}
  \script{rotation_models_example.py}
\end{figure}

Rotation has two main evolutionary effects: (\emph{i}) mixing can change the core
size directly (see \citealt{heger:2000, maeder:00}), (\emph{ii}) by
inflating the equatorial region, rotation changes the temperature and
opacity structure, and therefore the line-driving of the wind
\cite[e.g.,][]{muller:2014, gagnier:2019}, affecting the
rate of recession of the convective core \citep[e.g.,][]{renzo:2017,
  renzo:2020ppi_conv}. Moreover, rotation can have a dynamical effect,
resulting in mass loss through the combination of centrifugal forces
and radiative pressure ($\Gamma-\Omega$ limit, \citealt{langer:1998}).
One-dimensional stellar evolution codes commonly assume that rotation
increases the total mass loss rate \citep[e.g.,][]{langer:1998,
  heger:2000} though this may not always be true throughout the
evolution \citep[e.g.,][]{gagnier:2019}.

\Figref{fig:rotation_models_example} shows the gravitational binding
energy profile of the single, non-rotating
$18\,M_\odot$ star, compared to single stars of the
same mass and varying initial $\omega/\omega_{\rm crit}$.  For
$\omega/\omega_\mathrm{crit}\lesssim 0.5$, corresponding to a generous
upper-bound for the typical birth rotation rate of single massive
stars \citep[e.g.,][]{ramirez-agudelo:2015}, the effect is modest but
non-negligible. For more extreme initial rotation rates (achievable
during RLOF), the ratio of the He core mass to total mass is
significantly changed by rotational mixing, which can result in larger
binding energy differences than changing the CEB region at fixed core
mass.

\begin{figure}[!htbp]
  \centering
  \includegraphics[width=0.5\textwidth]{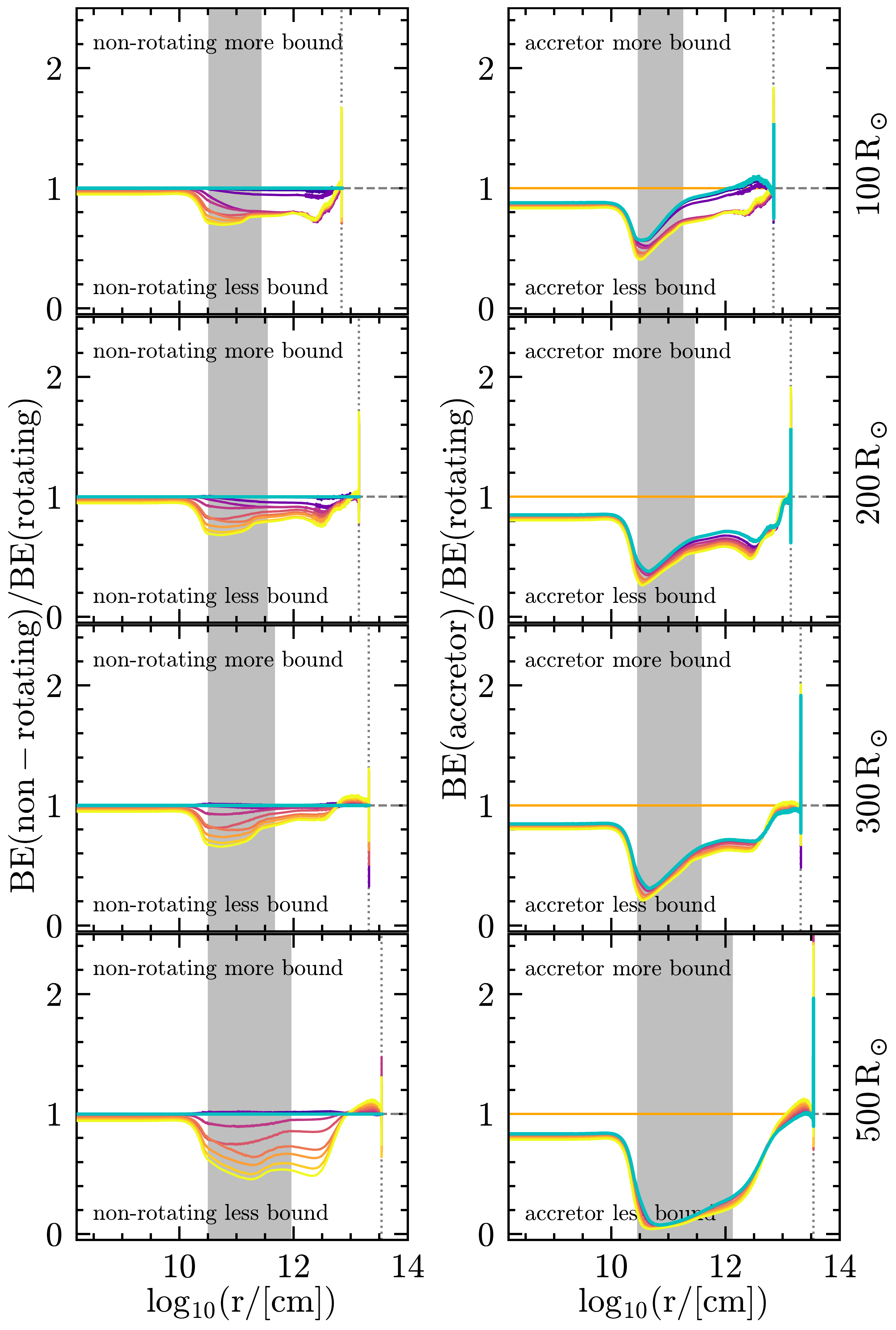}
  \caption{Ratio of binding energy of a non-rotating
      $18\,M_\odot$ model (left) or of our $15\rightarrow18\,M_\odot$
      accretor (right) divided by the binding energy of the rotating
      models of \Figref{fig:rotation_models_example}. The binding
      energy includes the internal energy ($\alpha_\mathrm{th}=1$),
      the gray bands highlight the CEB region of the models in the
      numerator. Lighter colors indicate ratios to an initially
      faster-rotating single star, see also legend in
      \Figref{fig:rotation_models_example}.}
  \label{fig:accretors_rotators_single}
  \script{accretors_rotators_single.py}
\end{figure}

\Figref{fig:accretors_rotators_single} shows the
  ratio of the binding energy (cf.~\Figref{fig:grid_ratios}) of a
  reference model divided the binding energy of the rotating models of
  \Figref{fig:rotation_models_example}. The left column uses as
  reference model for the numerator the non-rotating single
  $18\,M_\odot$, while the right column uses our
  $15\rightarrow18\,M_\odot$ accretor. The ordering of colors shows
  that the faster the initial rotation, the larger its structural
  effect on the star. However, single star models, regardless of their
  initial rotation rate, are more similar to each other than any
  single rotating star is to the accretor: in each row, the ratios in
  the left column are closer to one than the ration in the right
  column.  Moreover, the binding energy profiles of fast-rotating
  models (yellow) differ more than slow- and non-rotating models (blue
  and cyan) when compared to our accretor (i.e., their ratios are
  farther from one). Therefore, we do \emph{not} recommend the use of
  fast-rotating single stars to mimic the effect of mass accretion and
  rejuvenation.
For stars accreting through RLOF in a binary both
effects illustrated in \Figref{fig:toy_models_example} and
\Figref{fig:rotation_models_example} act simultaneously, although the
timing and amplitude of the impact of mixing and rotation can be
different than for single stars \citep[e.g.,][]{renzo:2021zoph}.
Future work should investigate how to include the
  effect in rapid population-synthesis, for example with
  a prescription for $\lambda_{CE}$ (see Appendix~\ref{sec:pop_synth_app}).

\section{Binding energy profiles}
\label{sec:BE}

\begin{figure*}[hbtp]
  \includegraphics[width=\textwidth]{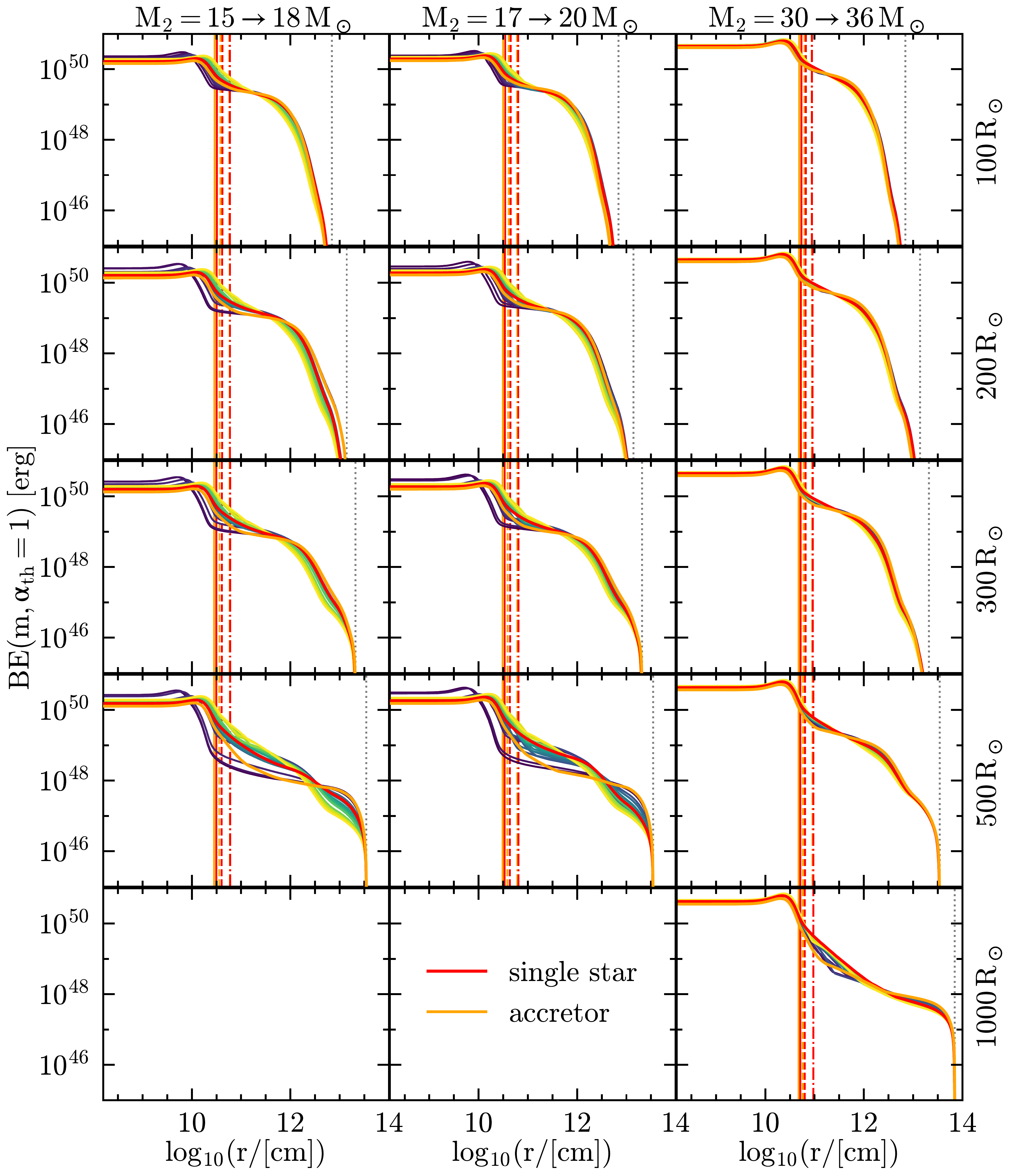}
  \caption{Binding energy profile at fixed photospheric radius $R$
    (right y-axis) as a function of radial coordinate $r$. We only
    show profiles with $\alpha_\mathrm{th}=1$, that is accounting for
    the internal energy content of the star. Orange, red, and other
    colors show respectively the accretor models, single stars of same
    post-RLOF total mass, and engineered models with varying CEB
    steepness (increasing CEB size from blue to
      yellow, cf.~\Figref{fig:TAMS_profiles}). Titles indicates the pre-RLOF and approximate
    post-RLOF accretor masses. The vertical colored lines mark the
    outer edge of the helium cores of the accretor and single star,
    that is the outermost location where $Y>0.1$ and $X<0.01$ (solid
    lines), or $X<0.1$ (dashed), or $X<0.2$ (dot-dashed). The dotted
    gray lines mark the total radius $R$ of these models.}
  \label{fig:BE_profiles}
  \script{BE_profiles.py}
\end{figure*}

In \Figref{fig:BE_profiles}, we show the binding energy of our
accretor models (solid lines,
including the internal energy, i.e.\ $\alpha_\mathrm{th}=1$ in
\Eqref{eq:BE}),
% and gravitational binding energy (dashed
% lines, $\alpha_\mathrm{th}=0$)
single stars with initial mass roughly equal
to the corresponding accretor's post-RLOF mass, and our engineered models (see also
\Figref{fig:lambda_grid} for the $\lambda_\mathrm{CE}$ profile defined in Appendix~\ref{sec:pop_synth_app}). The two
lowest mass accretors (left and central column) do not expand to
$R=1000\, R_\odot$ before carbon depletion. Generally speaking, the
accretors (orange) have lower binding energies than corresponding
single stars (red), and their profiles are qualitatively closer to the
engineered models with the steepest core (darker curves), although
local deviations from this trend can occur for some $r$.

\newpage
\section{Comparison with same core mass}
\label{sec:same_core}

\begin{figure*}[htbp]
  \centering
  \includegraphics[width=\textwidth]{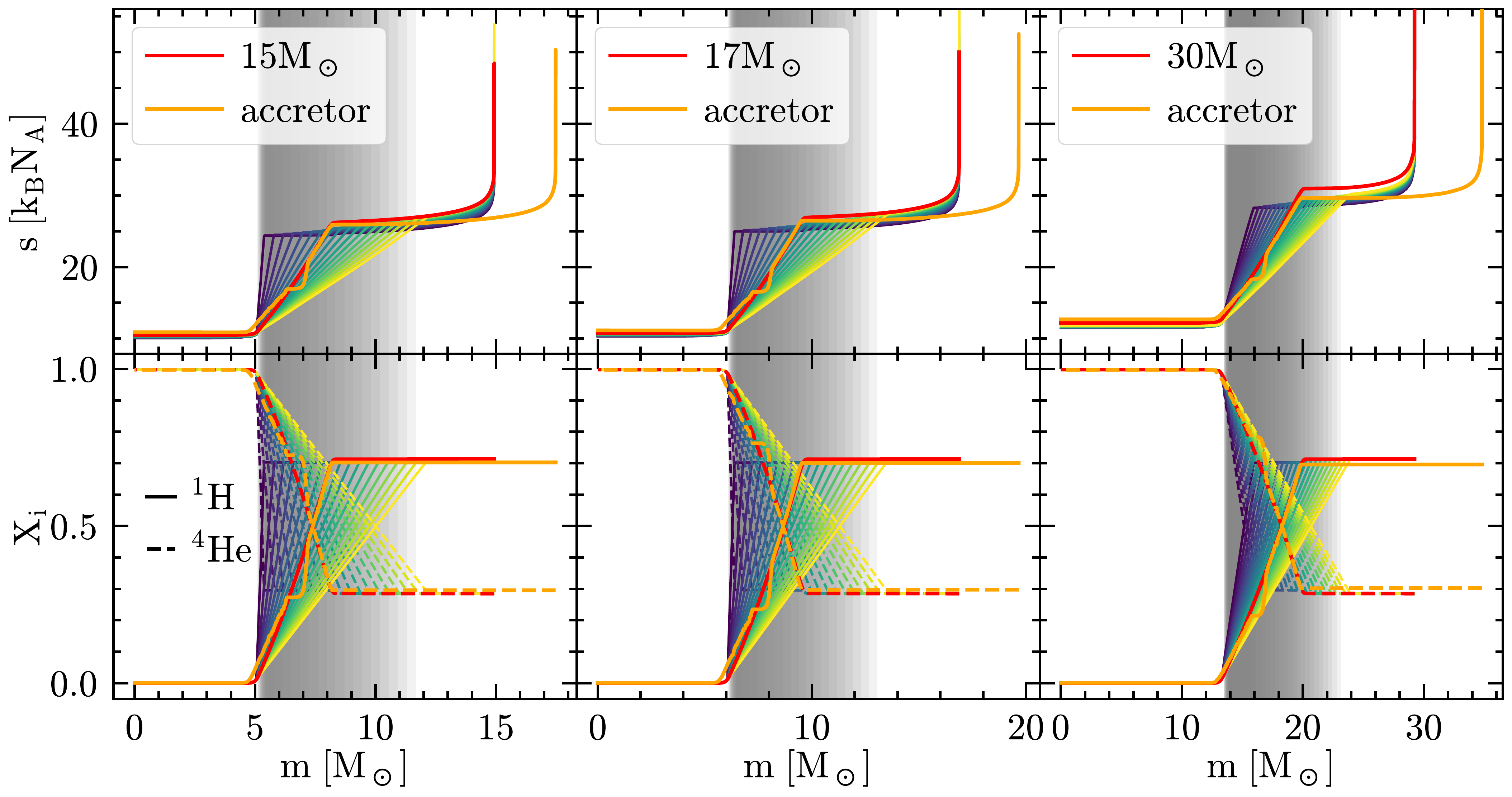}
  \caption{Specific entropy (top row), H (bottom row, solid lines),
    and He (bottom row, dashed lines) profiles for non-rotating single
    stars (red), accretors (orange), and ``engineered'' models of the
    same total mass as the ZAMS mass of the accretors. The overlapping
    gray bands emphasize the CEB region, increasing
      in size from blue to yellow in the engineered models.}
  \label{fig:TAMS_profiles_same_initial_mass}
  \script{TAMS_profiles_same_initial_mass.py}
\end{figure*}

\Figref{fig:TAMS_profiles} compares our accretor models to stars of
the same total \emph{post-RLOF} mass. However, it is not obvious that models
of the same total mass are the most relevant comparison: for instance,
the (helium or carbon-oxygen) core mass is often used to determine the
final compact object \citep[e.g.,][]{fryer:2012, farmer:2019,
  patton:2020, renzo:2022, fryer:2022}, and comparing models of
roughly the same core mass might be more appropriate (but is sensitive
to the condition defining the core edge). We show in
\Figref{fig:TAMS_profiles_same_initial_mass} a comparison of our
accretors with models of the same total \emph{initial} mass, which
constitute the extreme opposite comparison point.

\section{Common envelope $\lambda_\mathrm{CE}$}
\label{sec:pop_synth_app}

\begin{figure*}[htbp]
  \centering
  \includegraphics[width=\textwidth]{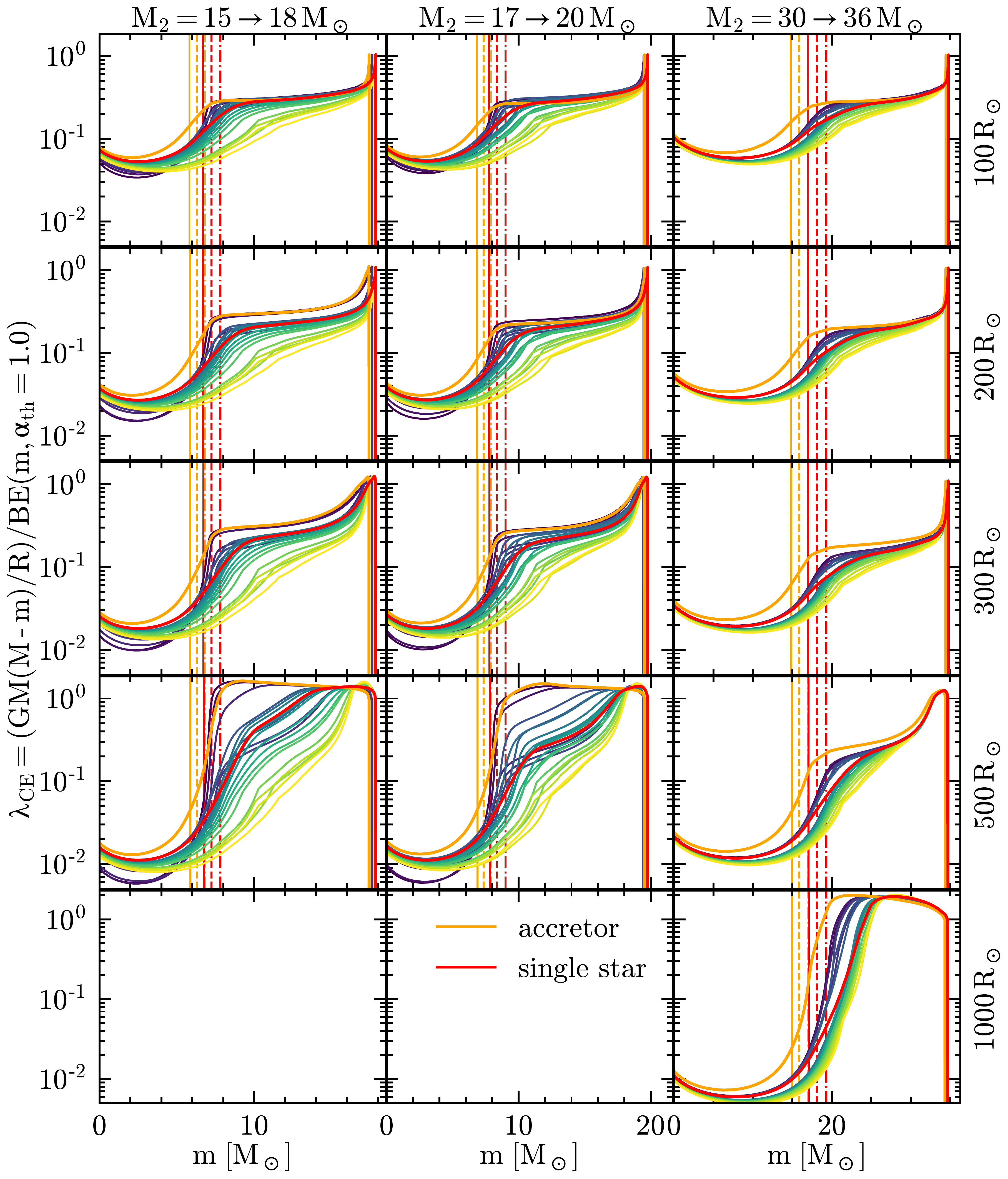}
  \caption{Profile of the binding energy parameter
    $\lambda_\mathrm{CE}$ as a function of mass coordinate for
    accretors (orange), single stars (red), and our engineered stars
    (other colors) at selected total radii. The vertical lines mark
    the outer edge of the helium cores of the accretor and single
    star, that is the outermost location where $Y>0.1$ and $X<0.01$
    (solid lines), or $X<0.1$ (dashed), or $X<0.2$
    (dot-dashed). The CEB size of engineered models
      increases from blue to yellow.}
  \label{fig:lambda_grid}
  \script{lambda_grid.py}
\end{figure*}

\cite{dekool:1990} introduced a binding energy parameter
$\lambda_\mathrm{CE}$ to account for the internal structure of the
stars when calculating the post-CE orbit using energy conservation:
\begin{equation}
  \label{eq:lambda}
  \lambda_\mathrm{CE} \equiv \lambda_\mathrm{CE}(m) = (GM(M-m)/R)/BE(m, \alpha_\mathrm{th}=1.0) \ \ ,
\end{equation}
where again the Lagrangian mass coordinate $m$ can be interpreted as a
variable core mass \citep[see also][]{demarco:11, ivanova:2013}. While
\cite{dekool:1990} implicitly used
  $\alpha_\mathrm{th}=0$, we calculate $\lambda_\mathrm{CE}$ with
  $\alpha_\mathrm{th}=1.0$ (including recombination energy), which provides a best case scenario for
  the ejection of the CE by harvesting the entire internal energy
  available in the gas. We show
in \Figref{fig:lambda_grid} the $\lambda_\mathrm{CE}$ profiles for our models.

\end{document}